\documentclass{amsart}
\usepackage{amsthm}
\usepackage{amssymb}
\usepackage{mathpartir}
\usepackage{graphicx}
\usepackage{multicol}
\usepackage{url}
\usepackage{preamble}
\usepackage{macros}
\usepackage{local-tikz}
\usepackage{categories}

\usepackage[natbib,style=alphabetic,useprefix=true,maxbibnames=99]{biblatex}
\newtheorem{theorem}{Theorem}
\newtheorem{therm}[theorem]{Theorem}
\newtheorem{lemma}[theorem]{Lemma}
\newtheorem{proposition}[theorem]{Proposition}
\newtheorem{remark}[theorem]{Remark}
\newtheorem{definition}[theorem]{Definition}
\newtheorem{notation}[theorem]{Notation}
\newtheorem{corollary}[theorem]{Corollary}
\newtheorem{example}[theorem]{Example}

\theoremstyle{remarkstyle}
\newtheorem{construction}[therm]{Construction}

\title{Controlling unfolding in type theory}

\author[Gratzer]{Daniel Gratzer}
\email{gratzer@cs.au.dk}
\author[Sterling]{Jonathan Sterling}
\email{js2878@cl.cam.ac.uk}
\author[Angiuli]{Carlo Angiuli}
\email{cangiuli@iu.edu}
\author[Coquand]{Thierry Coquand}
\email{coquand@chalmers.se}
\author[Birkedal]{Lars Birkedal}
\email{birkedal@cs.au.dk}
\date{\today}

\begin{document}

\begin{abstract}
  We present a new way to control the unfolding of definitions in dependent type theory.
  Traditionally, proof assistants require users to fix whether each definition will or will not be
  unfolded in the remainder of a development; unfolding definitions is often necessary in order to
  reason about them, but an excess of unfolding can result in brittle proofs and intractably large
  proof goals.
  In our system, definitions are by default not unfolded, but users can selectively unfold them in a
  local manner. We justify our mechanism by means of elaboration to a core theory with
  \emph{extension types}---a connective first introduced in the context of homotopy type
  theory---and by establishing a normalization theorem for our core calculus. We have implemented
  controlled unfolding in the \cooltt{} proof assistant, inspiring an independent implementation in
  Agda.
\end{abstract}

\maketitle

\section{Introduction}
\label{sec:intro}

In dependent type theory, terms are type checked modulo definitional equality,
a congruence generated by $\alpha$-, $\beta$-, and $\eta$-laws, as well as
unfolding of definitions. Unfolding definitions is to some extent a
convenience that allows type checkers to silently discharge many proof
obligations, \eg{} a list of length $1+1$ is without further annotation also a
list of length $2$. It is by no means the case, however, that we always want a
given definition to unfold:

\begin{itemize}
  \item \emph{Modularity}: Dependent types are famously sensitive to the smallest changes to
  definitions, such as whether $\prn{+}$ recurs on its first or its second argument. If we plan to
  change a definition in the future, it may be desirable to avoid exposing its implementation to the
  type checker.

  \item \emph{Usability}: While unfolding may simplify proof states, it also has the potential to
  complicate them, resulting in unreadable subgoals, error messages, \etc{} A user may find that
  certain definitions are likely to be problematic in this way, and thus opt not to unfold them.
\end{itemize}

Many proof assistants accordingly have implementation-level support for marking definitions
\emph{opaque} (unable to be unfolded), including Agda's \texttt{abstract} \citep{agda} and Coq's
\texttt{Qed}~\citep{coq}.
But unfolding definitions is not merely a matter of convenience: to reason about a function,
we must unfold it. For example, if we make the definition of $\prn{+}$ opaque, then
$\prn{+}$ is indistinguishable from a variable of type $\mathbb{N}\to\mathbb{N}\to\mathbb{N}$ and so
cannot be shown to be commutative, satisfy $1+1=2$, \etc{}

In practice, proof assistants resolve this contradiction by adopting an intermediate stance:
definitions are \emph{transparent} (unfolded during type checking) by default, but users are given
some control over their unfolding. Coq provides conversion tactics (\texttt{cbv}, \texttt{simpl},
\etc{}) for applying definitional equalities, each of which accepts a list of
definitions to unfold; its \texttt{Opaque} and \texttt{Transparent} commands toggle the default
unfolding behavior of a transparent definition; and the \textsc{SSReflect} tactic language natively
supports a ``\texttt{lock}ing'' idiom for controlling when definitions unfold \citep{gonthier:2009}.
Agda allows users to group multiple definitions into a single \texttt{abstract} block, inside of
which those definitions are transparent and outside of which they are opaque; this allows users to
define a function, prove all lemmas that depend on the function's definition, and then
irreversibly make the function and lemmas opaque.

These mechanisms for controlling unfolding pose interesting tradeoffs for users: which definitions
should be transparent, and which should be opaque? Transparency is in some cases necessary and in
many cases convenient, but is problematic both from an engineering perspective---because any edit to
a transparent definition can break the well-typedness of any number of its use sites---and from a
performance perspective---because checking definitional equality of type indices often requires
unfolding nested definitions into large normal forms.

In addition, the behavior of these mechanisms is more subtle than it may at first appear. In Agda,
definitions within \texttt{abstract} blocks are transparent to other definitions in the same block,
but opaque to the \emph{types} of those definitions; without such a stipulation, those types may
cease to be well-formed when the earlier definition is made opaque. Furthermore, \texttt{abstract}
blocks are anti-modular, requiring users to anticipate all future lemmas about definitions in a
block.%
\footnote{Indeed, the Agda standard library \citep{agda-stdlib} currently uses \texttt{abstract} only once.}
Coq's conversion tactics are more flexible than Agda's \texttt{abstract}
blocks, but being tactics, their behavior can be harder to predict. The \texttt{lock} idiom in
\textsc{SSReflect} is more predictable because it creates opaque definitions, but comes in four
different variations to simplify its use in practice.

\subsection{Contributions}
We propose a mechanism for fine-grained control over the unfolding of definitions in dependent
type theory. We introduce language-level primitives for \emph{controlled unfolding} that are
elaborated into a core calculus with \emph{extension types}, a connective first introduced by
\citet{riehl:2017}. We justify our elaboration algorithm by establishing a normalization theorem
(and hence the decidability of type checking and injectivity of type constructors) for our core
calculus, and we have implemented our system for controlled unfolding in the experimental \cooltt{}
proof assistant \citep{cooltt}.

Definitions in our framework are opaque by default, but can be selectively and locally unfolded as
if they were transparent. Our system is finer-grained and more modular than Agda's \texttt{abstract}
blocks: we need not collect all lemmas that unfold a given definition into a single block, making
our mechanism better suited to libraries. Our primitives have more predictable meaning and
performance than Coq's unfolding tactics%
\footnote{\url{https://github.com/coq/coq/blob/V8.16.0/theories/ssr/ssreflect.v/\#L388}}
because they are implemented by straightforward elaboration
into a core Martin-L{\"o}f type theory extended with new types and declaration forms.

In particular, we refine earlier approaches to representing definitions within type
theory~\citep{milner:1997,harper:2000,dreyer:2003,sterling:modules:2021} in order to more faithfully
represent definitions as they are actually used in practice: as neither fully opaque or transparent
but instead a mix of the two. Drawing inspiration from cubical type
theory~\citep{cohen:2017,angiuli:2018,angiuli:2021}, we extend \MLTT{} with proof-irrelevant
proposition symbols $p$, dependent products $\brc{p}\,A$ over those propositions, and extension
types $\Ext{A}{p}{a}$, the subtype of $A$ consisting of the elements of $A$ that definitionally
equal $a$ under the assumption that $p$ is true. For readers familiar with cubical type theory,
extension types are similar to path types $\prn{\Con{Path}~A~a_0~a_1}$, which classify functions out
of an abstract interval $\mathbb{I}$ that are definitionally equal to $a_0$ and $a_1$ when evaluated
at the interval's endpoints $0,1:\mathbb{I}$.

Encoding definitions through particular types confers a number of benefits. For instance, our
mechanism for definitions and unfolding are automatically invariant under definitional equivalence:
replacing one term by a definitionally-equal alternative cannot change the unfolding behavior of a
program. Furthermore, using extension types to encode definitions ensures our elaboration algorithm
is extremely modular and predictable: the rules for extension types are simple and, once grasped, it
becomes easy to predict the interactions between unfolding definitions and other features within the
language. This elaboration algorithm then serves as a \emph{reference} for the behavior of our
mechanism, against which other implementation strategies may be checked.

Like many elaboration algorithms for dependent type theory, executing our elaboration algorithm
requires deciding the equality of types in the core language. To show that our elaboration algorithm
can be implemented, we prove a \emph{normalization} theorem for our core calculus, characterizing
its definitional equivalence classes of types and terms and as a corollary establishing the
decidability of type checking. This is more subtle than it may appear: the heart of our
normalization proof amounts to correctly tracking when definitions are allowed to unfold as well as
when they should remain opaque. In the face of higher-order programs and dependent types, this is
quite difficult.

Another benefit to shifting from opaque definitions to extension types is their well-studied
metatheory. Specifically, we are able to adapt and extend Sterling's technique of synthetic Tait
computability / STC \citep{sterling:modules:2021,sterling:2021,sterling:phd} to prove normalization
for our core language.
Our proof is fully constructive, an improvement on the prior work of \citet{sterling:2021}; we have
also corrected an error in the handling of universes in an earlier revision of Sterling's doctoral
dissertation~\citep{sterling:phd} that was detected while preparing this paper.

\subsection{Outline}
In Section~\ref{sec:controlled-unfolding} we introduce our controlled unfolding primitives by way of
examples, and in Section~\ref{sec:informal-elaboration} we walk through how these examples are elaborated
into our core language of type theory with proposition symbols and extension types. In
Section~\ref{sec:elaboration} we present our elaboration algorithm, and in Section~\ref{sec:cooltt} we discuss our
implementation of the above in the \cooltt{} proof assistant. In Section~\ref{sec:normalization} we
establish normalization and its corollaries for our core calculus. We conclude with a discussion of
related work in Section~\ref{sec:related-work}.

\NewDocumentCommand{\VCons}{}{\mathbin{::}}
\NewDocumentCommand{\VNil}{}{{[\hspace{0.1em}]}}

\section{A surface language with controlled unfolding}
\label{sec:controlled-unfolding}

We begin by describing an Agda-like surface language for a dependent type
theory with controlled unfolding. In Section~\ref{sec:elaboration} we will give
precise meaning to this language by explaining how to elaborate it into our
core calculus; for now we proceed by example, introducing our new primitives
bit by bit.  Our examples will concern the inductively defined natural numbers
and their addition function:

\iblock{
  \mrow{\prn{+} : \Nat\to\Nat\to\Nat}
  \mrow{\Con{ze} + n = n}
  \mrow{{\Con{su}~m} + n = \Con{su}~\prn{m + n}}
}

\subsection{A simple dependency: length-indexed vectors}\label{sec:controlled-unfolding:top-level}

In our language, definitions such as $\prn{+}$ are opaque by default---they are not unfolded
automatically. To illustrate the need to \emph{selectively} unfold $\prn{+}$, consider the indexed
inductive type of length-indexed vectors with the following constructors:

\iblock{
  \mrow{\VNil{} : \Con{vec}~\Con{ze}~A}
  \mrow{\prn{\VCons} : A\to \Con{vec}~n~A\to \Con{vec}~\prn{\Con{su}~n}~A}
}

Suppose we attempt to define the \emph{append} operation on vectors by dependent pattern matching on
the first vector. Our goals would be as follows:

\iblock{
  \mrow{\prn{\oplus} : \Con{vec}~m~A\to \Con{vec}~n~A\to \Con{vec}\,\prn{m+n}~A}
  \mrow{\VNil{} \oplus v = \TypedHole{\Con{vec}~\prn{\Con{ze}+n}~A}}
  \mrow{\prn{a \VCons{} u} \oplus v = \TypedHole{\Con{vec}~\prn{\Con{su}~m+n}~A}}
}

As it stands, the goals above are in normal form and cannot be proved; however, we may indicate that
the definition of $\prn{+}$ should be unfolded within the definition of $\prn{\oplus}$ by adding the
following top-level \Kwd{unfolds} annotation:

\iblock{
  \mrow{\Spotlight{\prn{\oplus}~\Kwd{unfolds}~\prn{+}}}
  \mrow{
    \prn{\oplus} : \Con{vec}~m~A\to \Con{vec}~n~A\to \Con{vec}\,\prn{m+n}~A
  }
}

With our new declaration, the goals simplify:

\iblock{
  \mrow{\VNil{} \oplus v = \TypedHole{\Con{vec}~n~A}}
  \mrow{\prn{a \VCons{} u} \oplus v = \TypedHole{\Con{vec}~\prn{\Con{su}~\prn{m+n}}~A}}
}

The first goal is solved with $v$ itself; for the second goal, we begin by applying the
$\Con{vcons}$ constructor:

\iblock{
  \mrow{
    \prn{a \VCons{} u} \oplus v = a \VCons{} \TypedHole{\Con{vec}~\prn{m+n}~A}
  }
}

The remaining goal is just our induction hypothesis $u\oplus v$. All in all, we have:

\iblock{
  \mrow{\prn{\oplus}~\Kwd{unfolds}~\prn{+}}
  \mrow{\prn{\oplus} : \Con{vec}~m~A\to \Con{vec}~n~A\to \Con{vec}\,\prn{m+n}~A}
  \mrow{\VNil{} \oplus v = v}
  \mrow{\prn{a \VCons{} u} \oplus v = a \VCons{} \prn{u\oplus v}}
}

\subsection{Transitive unfolding}

Now suppose we want to prove that $\Con{map}$ distributes over $\prn{\oplus}$. In doing so we will
certainly need to unfold $\Con{map}$, but it turns out this will not be enough:

\iblock{
  \mrow{\Con{map} : \prn{A\to B} \to \Con{vec}~n~A \to \Con{vec}~n~B}
  \mrow{\Con{map}~f~\VNil{} = \VNil{}}
  \mrow{\Con{map}~f~\prn{a \VCons{} u} = f~a \VCons{} \Con{map}~f~u}

  \row

  \mrow{\MapAppend~\Kwd{unfolds}~\Con{map}}
  \mrow{
    \MapAppend : \prn{f:A\to B}~\prn{u:\Con{vec}~m~A}~\prn{v:\Con{vec}~n~A}
    \to \Con{map}~f~\prn{u\oplus v} \equiv \Con{map}~f~u \oplus \Con{map}~f~v
  }

  \mrow{
    \MapAppend~f~\VNil{}~v =
    \TypedHole{
      \Con{map}~f~\prn{\VNil{} \oplus v} \equiv \VNil{} \oplus \Con{map}~f~v
    }
  }

  \mrow{
    \MapAppend~f~\prn{a \VCons{} u}~v =
    \TypedHole{
      \Con{map}~f~\prn{a \VCons{} u} \oplus v \equiv
      \prn{f~a \VCons{} \Con{map}~f~u} \oplus \Con{map}~f~v
    }
  }
}

To make further progress we must also unfold $\prn{\oplus}$:

\iblock{
  \mrow{\MapAppend~\Kwd{unfolds}~\Con{map}; \Spotlight{\prn{\oplus}}}
  \mrow{
    \MapAppend : \prn{f:A\to B}~\prn{u:\Con{vec}~m~A}~\prn{v:\Con{vec}~n~A}
    \to \Con{map}~f~\prn{u\oplus v} \equiv \Con{map}~f~u \oplus \Con{map}~f~v
  }

  \mrow{
    \MapAppend~f~\VNil{}~v =
    \TypedHole{
      \Con{map}~f~v \equiv \Con{map}~f~v
    }
  }

  \mrow{
    \MapAppend~f~\prn{a \VCons{} u}~v =
    \TypedHole{
      f~a \VCons{} \Con{map}~f~\prn{u\oplus v} \equiv
      \prn{f~a \VCons{} \Con{map}~f~u} \oplus \Con{map}~f~v
    }
  }
}

In our language, unfolding $\prn{\oplus}$ has the side effect of \emph{also} unfolding $\prn{+}$: in
other words, unfolding is \emph{transitive}. To see why this is the case, observe that the unfolding
of $\prn{a \VCons{} u} \oplus v : \Con{vec}~\prn{{\Con{su}~m} + n}~A$, namely
$a \VCons{} \prn{u\oplus v} : \Con{vec}~\prn{\Con{su}~\prn{m + n}}~A$, would otherwise not be
well-typed. From an implementation perspective, one can think of the transitivity of unfolding as
necessary for \emph{subject reduction}. Having unfolded $\Con{map}$, $\prn{\oplus}$, and thus
$\prn{+}$, we complete our definition:

\iblock{
  \mrow{\Con{cong} : \prn{f:A\to B}\to a\equiv a'\to f~a\equiv f~a'}
  \mrow{\Con{cong}~f~\Con{refl} = \Con{refl}}
  \row
  \mrow{\MapAppend~\Kwd{unfolds}~\Con{map}; \prn{\oplus}}
  \mrow{
    \MapAppend : \prn{f:A\to B}~\prn{u:\Con{vec}~m~A}~\prn{v:\Con{vec}~n~A}
    \to \Con{map}~f~\prn{u\oplus v} \equiv \Con{map}~f~u \oplus \Con{map}~f~v
  }
  \mrow{\MapAppend~f~\VNil{}~v = \Con{refl}}
  \mrow{
    \MapAppend~f~\prn{a \VCons{} u}~v =
    \Con{cong}~\prn{f~a \VCons{}}~\prn{\MapAppend~f~u~v}
  }
}

\subsection{Recovering unconditionally transparent/opaque definitions}\label{sec:controlled-unfolding:abbreviation}

There are also times when we intend a given definition to be a fully transparent
\emph{abbreviation}, in the sense of being unfolded automatically whenever possible. We indicate
this with an $\Kwd{abbreviation}$ declaration:

\iblock{
  \mrow{\Spotlight{\Kwd{abbreviation}~\Con{singleton}}}
  \mrow{\Con{singleton} : A \to \Con{vec}~\prn{\Con{su}~\Con{ze}}~A}
  \mrow{\Con{singleton}~a = a \VCons{} \VNil{}}
}

Then the following lemma can be defined without any explicit unfolding:

\iblock{
  \mrow{\Con{abbrv\mhyphen{}example} : \Con{singleton}~5 \equiv \prn{5 \VCons{} \VNil{}}}
  \mrow{\Con{abbrv\mhyphen{}example} = \Con{refl}}
}

The meaning of the $\Kwd{abbreviation}$ keyword must account for unfolding constraints. For
instance, what would it mean to make $\MapAppend$ an abbreviation?

\iblock{
  \mrow{\Kwd{abbreviation}~\MapAppend}
  \mrow{\MapAppend~\Kwd{unfolds}~\Con{map}; \prn{\oplus}}
  \row \dots
}

We cannot unfold $\MapAppend$ in all contexts, because its definition is only well-typed
when $\Con{map}$ and $\prn{\oplus}$ are unfolded. The meaning of this declaration must, therefore,
be that $\MapAppend$ shall be unfolded \emph{just as soon as} $\Con{map}$ and $\prn{\oplus}$
are unfolded. In other words, $\Kwd{abbreviation}~\vartheta$ followed by
$\vartheta~\Kwd{unfolds}~\kappa_1;\ldots;\kappa_n$ means that unfolding $\vartheta$ is synonymous
with unfolding all of $\kappa_1;\ldots;\kappa_n$.

Conversely, we may intend a given definition \emph{never} to unfold, which we may indicate by a
corresponding $\Kwd{abstract}$ declaration. Because definitions in our system do not automatically unfold, the
force of $\Kwd{abstract}~\vartheta$ is simply to prohibit users from including $\vartheta$ in any
subsequent $\Kwd{unfolds}$ annotations.

\begin{remark}\label{rem:unfolding-sections}
A slight variation on our system can recover the behavior of Agda's \texttt{abstract} blocks by
\emph{limiting} the scope in which a definition $\vartheta$ can be unfolded; the transitivity of
unfolding dictates that any definition $\vartheta'$ that unfolds $\vartheta$ cannot itself be
unfolded once we leave that scope. We leave the details to future work.
\end{remark}

\subsection{Unfolding within the type}\label{sec:controlled-unfolding:within-type}

The effect of a $\vartheta~\Kwd{unfolds}~\kappa_1;\ldots;\kappa_n$ declaration
is to make $\kappa_1;\ldots\kappa_n$ unfold within the \emph{definition} of
$\vartheta$, but still not within its type; it will happen, however, that a
\emph{type} might not be expressible without some unfolding.  First we will show
how to accommodate this situation using only features we have introduced so
far, and then in Section~\ref{sec:controlled-unfolding:expression-level} we will
devise a more general and ergonomic solution.

Consider the left-unit law for $\prn{\oplus}$: in order to state that a vector $u$ is equal to the
vector $\VNil{}\oplus u$, we must contend with their differing types $\Con{vec}~n~A$ and
$\Con{vec}~\prn{\Con{ze}+n}~A$ respectively. One approach is to rewrite along the left-unit law for
$\mathbb{N}$; indeed, to state the \emph{right-unit} law for $\prn{\oplus}$ one must rewrite along
the right-unit law for $\mathbb{N}$. But here, because $\prn{+}$ computes on its first argument,
$\Con{vec}~n~A$ and $\Con{vec}~\prn{\Con{ze}+n}~A$ would be definitionally equal types if we could
unfold $\prn{+}$.

In order to formulate the left-unit law for $\prn{\oplus}$, we start by defining its \emph{type} as
an abbreviation that unfolds $\prn{+}$:

\NewDocumentCommand\AppendLeftUnit{}{\Con{{\oplus}\mhyphen{}left\mhyphen{}unit}}
\NewDocumentCommand\AppendLeftUnitType{}{\Con{{\oplus}\mhyphen{}left\mhyphen{}unit\mhyphen{}type}}

\NewDocumentCommand\AppendLeftUnitAlt{}{\Con{{\oplus}\mhyphen{}left\mhyphen{}unit'}}
\NewDocumentCommand\AppendLeftUnitAltType{}{\Con{{\oplus}\mhyphen{}left\mhyphen{}unit'\mhyphen{}type}}
\NewDocumentCommand\AppendLeftUnitAltBody{}{\Con{{\oplus}\mhyphen{}left\mhyphen{}unit'\mhyphen{}body}}

\iblock{
  \mrow{\Kwd{abbreviation}~\AppendLeftUnitType}
  \mrow{\AppendLeftUnitType~\Kwd{unfolds}~\prn{+}}
  \mrow{\AppendLeftUnitType : \Con{vec}~n~A\to \Con{Type}}
  \mrow{\AppendLeftUnitType~u = \VNil{}\oplus u \equiv u}
}

Now we may state the intended lemma using the type defined above:

\iblock{
  \mrow{\AppendLeftUnit : \prn{u : \Con{vec}~n~A} \to \AppendLeftUnitType~u}
  \mrow{\AppendLeftUnit~u = \TypedHole{\AppendLeftUnitType~u}}
}

Clearly we must unfold $\prn{+}$ and thus $\AppendLeftUnitType$ to
simplify our goal:

\iblock{
  \mrow{\AppendLeftUnit~\Kwd{unfolds}~\prn{+}}
  \mrow{\AppendLeftUnit : \prn{u : \Con{vec}~n~A} \to \AppendLeftUnitType~u}
  \mrow{\AppendLeftUnit~u = \TypedHole{\VNil{}\oplus u \equiv u}}
}

We complete the proof by unfolding $\prn{\oplus}$ itself, which transitively unfolds $\prn{+}$:

\iblock{
  \mrow{\AppendLeftUnit~\Kwd{unfolds}~\prn{\oplus}}
  \mrow{\AppendLeftUnit : \prn{u : \Con{vec}~n~A} \to \AppendLeftUnitType~u}
  \mrow{\AppendLeftUnit~u = \Con{refl}}
}

\subsection{Unfolding within subexpressions}\label{sec:controlled-unfolding:expression-level}

We have just demonstrated how to unfold definitions within the \emph{type} of a declaration by
defining that type as an additional declaration; using the same technique, we can introduce
unfoldings within \emph{any subexpression} by hoisting that subexpression to a top-level definition
with its own unfolding constraint.

\paragraph*{Unfolding within the type, revisited}
Rather than repeating the somewhat verbose pattern of Section~\ref{sec:controlled-unfolding:within-type},
we abstract it as a new language feature that is easily eliminated by elaboration. In particular, we
introduce a new \emph{expression} former $\Kwd{unfold}~\kappa~\Kwd{in}~M$ that can be placed in any
expression context. Let us replay the example from Section~\ref{sec:controlled-unfolding:within-type}, but
using $\Kwd{unfold}$ rather than an auxiliary definition:

\iblock{
  \mrow{
    \AppendLeftUnit : \prn{u : \Con{vec}~n~A} \to
    \Spotlight{\Kwd{unfold}~\prn{+}~\Kwd{in}}~\VNil{}\oplus u\equiv u
  }
  \mrow{
    \AppendLeftUnit~u =
    \TypedHole{
      \Kwd{unfold}~\prn{+}~\Kwd{in}~\VNil{}\oplus u\equiv u
    }
  }
}

The type $\Kwd{unfold}~\prn{+}~\Kwd{in}~\VNil{}\oplus u \equiv u$ is in
normal form; the only way to simplify it is to unfold $\prn{+}$. We could do
this with another inline $\Kwd{unfold}$ expression (see $\AppendLeftUnitAlt$ below), but here we will use a
top-level declaration:
\iblock{
  \mrow{\Spotlight{\AppendLeftUnit~\Kwd{unfolds}~\prn{+}}}
  \mrow{
    \AppendLeftUnit : \prn{u : \Con{vec}~n~A} \to
    \Kwd{unfold}~\prn{+}~\Kwd{in}~\VNil{}\oplus u\equiv u
  }
  \mrow{
    \AppendLeftUnit~u =
    \TypedHole{
      \VNil{}\oplus u\equiv u
    }
  }
}

By virtue of the above, the $\Kwd{unfold}$ expression in our hole has computed
away and we are left with $\TypedHole{\VNil{}\oplus u\equiv u}$ as
$\prn{\oplus}$ is still abstract in this scope. To make progress, we
\emph{strengthen} the declaration to unfold $\prn{\oplus}$ in addition to
$\prn{+}$:
\iblock{
  \mrow{\AppendLeftUnit~\Kwd{unfolds}~\Spotlight{\prn{\oplus}}}
  \mrow{\AppendLeftUnit : \prn{u : \Con{vec}~n~A} \to \Kwd{unfold}~\prn{+}~\Kwd{in}~\VNil{}\oplus u\equiv u}
  \mrow{\AppendLeftUnit~u = \Con{refl}}
}

The meaning of the code above is exactly as described in
Section~\ref{sec:controlled-unfolding:within-type}: the $\Kwd{unfold}$ scope
is elaborated to a new top-level $\Kwd{abbreviation}$ that unfolds
$\prn{+}$.

\paragraph*{Expression-level \vs top-level unfolding}
We noted in our definition of $\AppendLeftUnit$ above that we could have
replaced the top-level $\Kwd{unfolds}~\prn{\oplus}$ directive of $\AppendLeftUnit$ with the
new expression-level $\Kwd{unfold}~\prn{\oplus}~\Kwd{in}$ as follows:

\iblock{
  \mrow{\AppendLeftUnitAlt : \prn{u : \Con{vec}~n~A} \to \Kwd{unfold}~\prn{+}~\Kwd{in}~\VNil{}\oplus u\equiv u}
  \mrow{\AppendLeftUnitAlt~u = \Kwd{unfold}~\prn{\oplus}~\Kwd{in}~\Con{refl}}
}

The resulting definition of $\AppendLeftUnitAlt$ has slightly different behavior than
$\AppendLeftUnit$ above: whereas unfolding $\AppendLeftUnit$ causes $\prn{\oplus}$
to unfold transitively, we can unfold $\AppendLeftUnitAlt$ without unfolding
$\prn{\oplus}$---at the cost of $\Kwd{unfold}~\prn{\oplus}$ expressions appearing in our goal. This
more granular behavior may be desirable in some cases, and it is a strength of our language and its
elaborative semantics that the programmer can manipulate unfolding in such a fine-grained manner.

For completeness, we show the elaborated version of $\AppendLeftUnitAlt$ resulting from eliminating
expression-level unfolding from the definition. We defer a systematic discussion of this
transformation till Section~\ref{sec:elaboration}.

\iblock{
  \mrow{\Kwd{abbreviation}~\AppendLeftUnitAltType}
  \mrow{\AppendLeftUnitAltType~\Kwd{unfolds}~\prn{+}}
  \mrow{\AppendLeftUnitAltType : \Con{vec}~n~A\to \Con{Type}}
  \mrow{\AppendLeftUnitAltType~u = \VNil{}\oplus u \equiv u}
  \row
  \mrow{\Kwd{abbreviation}~\AppendLeftUnitAltBody}
  \mrow{\AppendLeftUnitAltBody~\Kwd{unfolds}~\prn{\oplus}}
  \mrow{\AppendLeftUnitAltBody : \prn{u : \Con{vec}~n~A}\to \AppendLeftUnitAltType~u}
  \mrow{\AppendLeftUnitAltBody~u = \Con{refl}}
  \row
  \mrow{\AppendLeftUnitAlt : \prn{u : \Con{vec}~n~A}\to \AppendLeftUnitAltType~u}
  \mrow{\AppendLeftUnitAlt~u = \AppendLeftUnitAltBody~u}
}

In our experience, expression-level unfolding seems more commonly useful for end users than
top-level unfolding; on the other hand, the clearest semantics for expression-level unfolding are
stated in terms of top-level unfolding! Because one of our goals is to provide an account of
unfolding that admits a reliable and precise mental model for programmers, it is desirable to
include both top-level and expression-level unfolding in the surface language.

\section{Controlling unfolding with extension types}
\label{sec:informal-elaboration}

Having introduced our new surface language constructs for controlled unfolding in
Section~\ref{sec:controlled-unfolding}, we now describe how to elaborate these constructs into our
dependently-typed core calculus. Again we proceed by example, deferring our formal descriptions of
the elaboration algorithm to Section~\ref{sec:elaboration}.

\subsection{A core calculus with proposition symbols}

Our core calculus parameterizes intensional Martin-L{\"o}f type theory (\MLTT)
\citep{martin-lof:1973} by a bounded meet semilattice of \emph{proposition symbols} $p\in\PP$, and
adjoins to the type theory a new form of context extension and two new type formers $\brc{p}\, A$
and $\Ext{A}{p}{M}$ involving proposition symbols:

\begin{grammar}
  contexts & \Gamma & \dots \GrmSep{} \Gamma,p
  \\
  types & A & \dots \GrmSep{} \brc{p}\, A \GrmSep{} \Ext{A}{p}{M}
\end{grammar}

The bounded meet semilattice structure on $\PP$ closes proposition symbols under conjunction $\land$
and the true proposition $\top$, thereby partially ordering $\PP$ by entailment $p \leq q$ (``$p$
entails $q$'') satisfying the usual principles of propositional logic. We say $p$ is \emph{true} if
$\top$ entails $p$; the context extension $\Gamma,p$ hypothesizes that $p$ is true.

\begin{remark}
Our proposition symbols are much more restricted than, and should not be confused with, other
notions of proposition in type theory such as h-propositions \cite[\S3.3]{hottbook} or strict
propositions \citep{gilbert:2019}. In particular, unlike types, our proposition symbols have no
associated proof terms.
\end{remark}

The type $\brc{p}\, A$ is the dependent product ``$\brc{\_ : p}\to A$'', \ie{}, $\brc{p}\, A$ is
well-formed when $A$ is a type under the hypothesis that $p$ is true, and $f : \brc{p}\, A$ when,
given that $p$ is true, we may conclude $f : A$. The \emph{extension type} $\Ext{A}{p}{a_p}$ is
well-formed when $A$ is a type and $a_p : \brc{p}\, A$; its elements $a : \Ext{A}{p}{a_p}$ are terms
$a : A$ satisfying the side condition that when $p$ is true, we have $a = a_p : A$. We provide
inference rules for the core calculus, including these connectives, in
Section~\ref{sec:elaboration:core-calculus}.

\subsection{Elaborating controlled unfolding to our core calculus}

Our surface language extends a generic surface language for dependent type theory with a new
expression former $\Kwd{unfold}$ and several new declaration forms:
$\vartheta~\Kwd{unfolds}~\kappa_1;\dots;\kappa_n$ for controlled unfolding,
$\Kwd{abbreviation}~\vartheta$ for transparent definitions, and $\Kwd{abstract}~\vartheta$ for
opaque definitions. Elaboration transforms these surface-language declarations into core-language
\emph{signatures}, \ie{} sequences of declarations over our core calculus of \MLTT{} with
proposition symbols.

Our signatures include the following declaration forms:
\begin{itemize}
\item $\Kwd{prop}~p \leq q$ introduces a fresh proposition symbol $p$ such that $p$ entails
  $q\in\PP$;
\item $\Kwd{prop}~p = q$ defines the proposition symbol $p$ to be an abbreviation for $q\in\PP$;
\item $\Kwd{const}~\vartheta : A$ introduces a constant $\vartheta$ of type $A$.
\end{itemize}

We now revisit our examples from Section~\ref{sec:controlled-unfolding}, illustrating how they are
elaborated into our core calculus:

\NewDocumentCommand\UnfTok{m}{\Upsilon\Sub{#1}}
\NewDocumentCommand\Defn{m}{\delta\Sub{#1}}

\subsection*{Plain definitions}
Recall our unadorned definition of $\prn{+}$ from Section~\ref{sec:controlled-unfolding}:

\iblock{
  \mrow{\prn{+} : \Nat\to\Nat\to\Nat}
  \mrow{\Con{ze} + n = n}
  \mrow{{\Con{su}~m} + n = \Con{su}~\prn{m + n}}
}

We elaborate $\prn{+}$ into a sequence of declarations: first, we introduce a new proposition symbol
$\UnfTok{+}$ corresponding to the proposition that ``$\prn{+}$ unfolds.'' Next, we introduce a new
definition $\Defn{+}:\Nat\to\Nat\to\Nat$ satisfying the defining clauses of $\prn{+}$ above, under
the (trivial) assumption of $\top$; finally, we introduce a new constant $\prn{+}$ involving the
extension type of $\Defn{+}$ along $\UnfTok{+}$.

\iblock{
  \mrow{\Kwd{prop}~\UnfTok{+}\leq\top}
  \row
  \mrow{\Defn{+} : \brc{\top}\,\prn{m\,n : \Nat}\to \Nat}
  \mrow{\Defn{+}~\Con{ze}~n = n}
  \mrow{\Defn{+}~\prn{\Con{su}~m}~n = \Con{su}~\prn{\Defn{+}~m~n}}
  \row
  \mrow{\Kwd{const}~\prn{+} : \Ext{\Nat\to\Nat\to\Nat}{\UnfTok{+}}{\Defn{+}}}
}

\begin{remark}
  In a serious implementation, it would be simple to induce $\Defn{+}$
  to be pretty-printed as $\prn{+}$ in user-facing displays such as goals
  and error messages.
\end{remark}

\subsection*{Top-level unfolding}
To understand why we have elaborated $\prn{+}$ in this way, let us examine how to elaborate
top-level unfolding declarations (Section~\ref{sec:controlled-unfolding:top-level}):

\iblock{
  \mrow{\prn{\oplus}~\Kwd{unfolds}~\prn{+}}
  \mrow{\prn{\oplus} : \Con{vec}~m~A\to \Con{vec}~n~A\to \Con{vec}\,\prn{m+n}~A}
  \mrow{\VNil{} \oplus v = v}
  \mrow{\prn{a \VCons{} u} \oplus v = a \VCons{} \prn{u\oplus v}}
}

To elaborate $\prn{\oplus}~\Kwd{unfolds}~\prn{+}$, we define the proposition symbol
$\UnfTok{\oplus}$ to entail $\UnfTok{+}$, capturing the idea that unfolding $\prn{\oplus}$ always
causes $\prn{+}$ to unfold; in order to cause $\prn{+}$ to unfold in the body of $\prn{\oplus}$, we
assume $\UnfTok{+}$ in the definition of $\Defn{\oplus}$. In full, we elaborate the definition of
$\prn{\oplus}$ as follows:

\iblock{
  \mrow{\Kwd{prop}~\UnfTok{\oplus}\leq\UnfTok{+}}
  \row
  \mrow{\Defn{\oplus} : \brc{\UnfTok{+}} \, \prn{u:\Con{vec}~m~A}\,\prn{v:\Con{vec}~n~A}\to \Con{vec}\,\prn{m+n}~A}
  \mrow{\Defn{\oplus}~\VNil{}~v = v}
  \mrow{\Defn{\oplus}~\prn{a \VCons{} u}~v = a \VCons{} \prn{\Defn{\oplus}~u~v}}
  \row
  \mrow{
    \Kwd{const}~\prn{\oplus} :
    \Ext{
      \Con{vec}~m~A\to \Con{vec}~n~A\to \Con{vec}\,\prn{m+n}~A
    }{\UnfTok{\oplus}}{\Defn{\oplus}}
  }
}

Observe that the definition of $\Defn{\oplus}$ is well-typed because $\UnfTok{+}$ is true in its
scope: thus the extension type of
$\prn{+}$ causes $\Con{ze}+n$ to be definitionally equal to $\Defn{+}~\Con{ze}~n$, which in turn is
defined to be $n$. The constraint $\UnfTok{\oplus}\hookrightarrow\Defn{\oplus}$ is well-typed
because $\UnfTok{\oplus}$ entails $\UnfTok{+}$.

If a definition $\vartheta$ unfolds multiple definitions $\kappa_1;\dots;\kappa_n$, we define
$\UnfTok{\vartheta}$ to entail (and define $\Defn{\vartheta}$ to assume) the conjunction
$\UnfTok{\kappa_1}\land\dots\land\UnfTok{\kappa_n}$; if a definition $\vartheta$ unfolds no
definitions, then $\UnfTok{\vartheta}$ entails (and $\Defn{\vartheta}$ assumes) $\top$, as in our
$\prn{+}$ example.

\subsection*{Abbreviations}

To elaborate the combination of the declarations $\Kwd{abbreviation}~\vartheta$ and
$\vartheta~\Kwd{unfolds}~\kappa_1;\dots;\kappa_n$ we define $\UnfTok{\vartheta}$ to \emph{equal} the
conjunction $\UnfTok{\kappa_1}\land\dots\land\UnfTok{\kappa_n}$. For example, consider
the following code from Section~\ref{sec:controlled-unfolding:abbreviation}:

\iblock{
  \mrow{\Kwd{abbreviation}~\MapAppend}
  \mrow{\MapAppend~\Kwd{unfolds}~\Con{map}; \prn{\oplus}}
  \mrow{
    \MapAppend : \prn{f:A\to B}~\prn{u:\Con{vec}~m~A}~\prn{v:\Con{vec}~n~A}
    \to
    \Con{map}~f~\prn{u\oplus v} \equiv \Con{map}~f~u \oplus \Con{map}~f~v
  }
  \mrow{\MapAppend~f~\VNil{}~v = \Con{refl}}
  \mrow{\MapAppend~f~\prn{a \VCons{} u}~v = \Con{cong}~\prn{\prn{f~a} \VCons{}-}~\prn{\MapAppend~f~u~v}}
}

Let us write $\mathfrak{C}$ for the following type:

\iblock{
  \mrow{
    \prn{f:A\to B}~\prn{u:\Con{vec}~m~A}~\prn{v:\Con{vec}~n~A}
    \to \Con{map}~f~\prn{u\oplus v} \equiv \Con{map}~f~u \oplus \Con{map}~f~v
  }
}

The above example is then elaborated as follows:

\iblock{
  \mrow{\Kwd{prop}~\UnfTok{\MapAppend} = \UnfTok{\Con{map}}\land \UnfTok{\oplus}}
  \row
  \mrow{\Defn{\MapAppend} : \brc{\UnfTok{\Con{map}}\land\UnfTok{\oplus}}\,\mathfrak{C}}
  \mrow{\Defn{\MapAppend}~f~\VNil{}~v = \Con{refl}}
  \mrow{\Defn{\MapAppend}~f~\prn{a \VCons{} u}~v = \Con{cong}~\prn{\prn{f~a} \VCons{} -}~\prn{\Defn{\MapAppend}~f~u~v}}
  \row
  \mrow{
    \Kwd{const}~
    \MapAppend :
    \lbrace \mathfrak{C} \mid \UnfTok{\MapAppend} \hookrightarrow \Defn{\MapAppend} \rbrace
  }
}

\subsection*{Expression-level unfolding}

The elaboration of the expression-level unfolding construct $\Kwd{unfold}~\kappa~\Kwd{in}~M$ to our
core calculus factors through the elaboration of expression-level unfolding to top-level unfolding
as described in Section~\ref{sec:controlled-unfolding:expression-level}; we return to this in
Section~\ref{sec:elaboration:elaboration}.

\section{The elaboration algorithm}
\label{sec:elaboration}

We now formally specify our mechanism for controlled unfolding by more precisely defining the
elaboration algorithm sketched in the previous section, starting with a precise definition of the
target of elaboration, our core calculus \TTProp{}.

\subsection{The core calculus \texorpdfstring{\TTProp{}}{TT\_P}}
\label{sec:elaboration:core-calculus}

Our core calculus \TTProp{} is intensional Martin-L{\"o}f type theory (\MLTT) \citep{martin-lof:1973}
with dependent sums and products, a Tarski universe, \etc{}, extended with (1) a
collection of proof-irrelevant proposition symbols, (2) dependent products over propositions, and
(3) extension types for those propositions~\citep{riehl:2017}.

\begin{remark}
  We treat the features of \MLTT{} and of our surface language somewhat generically; our elaboration
  algorithm can be applied on top of an existing bidirectional elaboration algorithm for type theory,
  \eg those described by \citet{dagand:phd,gratzer:2019}, which may separately account for features
  such as implicit arguments or dependent pattern matching.
\end{remark}

In fact, \TTProp{} is actually a family of type theories parameterized by a bounded meet semilattice
$(\PP,\top,\land)$ whose underlying set $\PP$ is the set of proposition symbols of \TTProp{}; the
semilattice structure on $\PP$ axiomatizes the conjunctive fragment of propositional logic with
$\land$ as conjunction, $\top$ as the true proposition, and $\leq$ as entailment (where $p\leq q$
is defined as $p\land q = p$), subject to the usual logical principles such as $p\land q \leq p$ and
$p\land q \leq q$ and $p \leq \top$.

\begin{remark}
  \label{rem:elaboration:functorial}
  The judgments of \TTProp{} are functorial in the choice of $\PP$, in the sense that given any
  homomorphism $\Mor[f]{\PP}{\PP'}$ of bounded meet semilattices and any type or term in \TTProp{}
  over $\PP$, we have an induced type/term in \TTPropp{\PP'} over $\PP'$. In particular, we will use
  the fact that judgments of \TTProp{} are stable under adjoining new proposition symbols to $\PP$.
\end{remark}

The language \TTProp{} augments ordinary \MLTT{} with a new judgment $\IsTrue{p}$ (for
$p\in\PP$) and the corresponding context extension $\RCx{\Gamma}{p}$ (for $p\in\PP$). The judgment
$\IsTrue{p}$ states that the proposition $p$ is true in context $\Gamma$, \ie{}, the conjunction of
the propositional hypotheses in $\Gamma$ entails $p$ while $\RCx{\Gamma}{p}$ extends $\Gamma$
with the hypothesis that $p$ is true.

\begin{mathparpagebreakable}
  \inferrule{
    \IsCx{\Gamma}
    \\
    p\in\PP
  }{
    \IsCx{\RCx{\Gamma}{p}}
  }
  \and
  \inferrule{
    p\in\PP
  }{
    \IsTrue[\RCx{\Gamma}{p}]{p}
  }
  \and
  \inferrule{
    \RCx{\Gamma}{p} \vdash {\mathcal{J}}
    \\
    \IsTrue{p}
  }{
    \Gamma \vdash \mathcal{J}
  }
  \\
  \inferrule{
  }{
    \IsTrue{\top}
  }
  \and
  \inferrule{
    \IsTrue{p}
    \\
    \IsTrue{q}
  }{
    \IsTrue{p\land q}
  }
  \and
  \inferrule{
    \IsTrue{p}
    \\
    p\leq q
  }{
    \IsTrue{q}
  }
\end{mathparpagebreakable}

The dependent product $\brc{p}\,A$ is defined as an ordinary dependent product:

\begin{mathparpagebreakable}
  \inferrule{
    \IsTy[\RCx{\Gamma}{p}]{A}
  }{
    \IsTy{\brc{p}\,A}
  }
  \and
  \inferrule{
    \IsTm[\RCx{\Gamma}{p}]{M}{A}
  }{
    \IsTm{\gl{p}\,M}{\brc{p}\,A}
  }
  \and
  \inferrule{
    \IsTm{M}{\brc{p}\,A}
    \\
    \IsTrue{p}
  }{
    \IsTm{M \mathbin{@} p}{A}
  }
  \and
  \inferrule{
    \IsTm[\RCx{\Gamma}{p}]{M}{A}
    \\
    \IsTrue{p}
  }{
    \EqTm{\prn{\gl{p}\,M} \mathbin{@} p}{M}{A}
  }
  \and
  \inferrule{
    \IsTm{M}{\brc{p}\,A}
  }{
    \EqTm{M}{\gl{p}\,\prn{M \mathbin{@} p}}{\brc{p}\,A}
  }
\end{mathparpagebreakable}

The remaining feature of \TTProp{} is the extension type $\Ext{A}{p}{a\Sub{p}}$.
Given a proposition $p\in\PP$ and an element $a\Sub{p}$ of $A$ under the hypothesis $p$,
the elements of $\Ext{A}{p}{a\Sub{p}}$ correspond to elements of $A$ that equal $a\Sub{p}$
when $p$ holds.

\begin{mathparpagebreakable}
  \inferrule{
    \IsTy{A}
    \\
    \IsTm[\RCx{\Gamma}{p}]{a\Sub{p}}{A}
  }{
    \IsTy{\Ext{A}{p}{a\Sub{p}}}
  }
  \and
  \inferrule{
    \IsTm{a}{A}
    \\\\
    \IsTm[\RCx{\Gamma}{p}]{a\Sub{p}}{A}
    \\\\
    \EqTm[\RCx{\Gamma}{p}]{a}{a\Sub{p}}{A}
  }{
    \IsTm{\In_p\,a}{\Ext{A}{p}{a\Sub{p}}}
  }
  \and
  \inferrule{
    \IsTm{a}{\Ext{A}{p}{a\Sub{p}}}
  }{
    \IsTm{\Out_p\,a}{A}
  }
  \and
  \inferrule{
    \IsTm{a}{A}
  }{
    \EqTm{\Out_p\prn{\In_p\,a}}{a}{A}
  }
  \and
  \inferrule{
    \IsTm{a}{\Ext{A}{p}{a\Sub{p}}}
  }{
    \EqTm{\In_p\prn{\Out_p\,a}}{a}{\Ext{A}{p}{a\Sub{p}}}
  }
  \and
  \inferrule{
    \IsTrue{p}
    \\
    \IsTm{a}{\Ext{A}{p}{a\Sub{p}}}
  }{
    \EqTm{\Out_p\,a}{a\Sub{p}}{A}
  }
\end{mathparpagebreakable}

\subsection{Signatures over \texorpdfstring{\TTProp{}}{TT\_P}}
\label{sec:elaboration:signatures}

Our elaboration procedure takes as input a sequence of surface language definitions, and outputs a
well-formed \emph{signature}, a list of declarations over \TTProp{}.

\begin{grammar}
  sigs & \Sigma & \epsilon \GrmSep{} \Sigma, D
  \\
  decls & D & \Kwd{const}\, x : A
  \GrmSep{} \Kwd{prop}\, p \le q
  \GrmSep{} \Kwd{prop}\, p = q
\end{grammar}

\NewDocumentCommand{\SigOK}{mmm}{\vdash #1\,\mathit{sig} \longrightarrow #2, #3}

A signature is well-formed precisely when each declaration in $\Sigma$ is well-formed relative to
the earlier declarations in $\Sigma$. Our well-formedness judgment
$\SigOK{\Sigma}{\mathbb{P}}{\Gamma}$ computes from $\Sigma$ the \TTProp{} context $\Gamma$ and
proposition semilattice $\mathbb{P}$ specified by $\Sigma$'s $\Kwd{const}$ and $\Kwd{prop}$
declarations, respectively.

The rules for signature well-formedness are standard except for the $\Kwd{prop}\,p \le q$ and
$\Kwd{prop}\,p = q$ declarations, which extend $\mathbb{P}$ with a new element $p$ satisfying $p \le
q$ or $p = q$ respectively. Recalling that our core calculus \TTProp{} is really a family of type
theories parameterized by a semilattice $\mathbb{P}$, these declarations shift us between type
theories, \eg{}, from \TTProp{} to \TTPropp{\mathbb{Q}}, where $\mathbb{Q} = \mathbb{P}\brk{p \le q}$
is the minimal semilattice containing $\mathbb{P}$ and an element $p$ satisfying $p \le q$. This
shifting between theories is justified by Remark~\ref{rem:elaboration:functorial}.

\begin{mathpar}
  \inferrule{ }{
    \SigOK{\epsilon}{\brc{\top}}{\cdot}
  }
  \and
  \inferrule{
    \SigOK{\Sigma}{\mathbb{P}}{\Gamma}
    \\
    \Gamma \vdash_{\text{\TTProp}} A\,\mathit{type}
  }{
    \SigOK{\prn{\Sigma,\ \Kwd{const}\, x : A}}{\mathbb{P}}{\prn{\Gamma, x : A}}
  }
  \and
  \inferrule{
    \SigOK{\Sigma}{\mathbb{P}}{\Gamma}
    \\
    q\in\mathbb{P}
  }{
    \SigOK{
      \prn{\Sigma,\ \Kwd{prop}\, p \le q}
    }{
      \mathbb{P}\brk{p \le q}
    }{
      \Gamma
    }
    \\
    \SigOK{
      \prn{\Sigma,\ \Kwd{prop}\, p = q}
    }{
      \mathbb{P}\brk{p = q}
    }{
      \Gamma
    }
  }
\end{mathpar}

\subsection{Bidirectional elaboration}
\label{sec:elaboration:elaboration}

\NewDocumentCommand{\Elab}{mmm}{#1 \vdash #2 \rightsquigarrow #3}
\NewDocumentCommand{\SynTm}{mmmm}{#1 \vdash #2 \Rightarrow #3 \rightsquigarrow #4}
\NewDocumentCommand{\ChkTm}{mmmm}{#1 \vdash #2 \Leftarrow #3 \rightsquigarrow #4}
\NewDocumentCommand{\ChkTp}{mmm}{#1 \vdash #2 \Leftarrow \mathit{type} \rightsquigarrow #3}

We adopt a bidirectional elaboration algorithm which mirrors bidirectional type-checking
algorithms~\citep{coquand:1996,pierce:2000}.
The top-level elaboration judgment $\Elab{\Sigma}{\vec{S}}{\Sigma'}$ takes as input the current
well-formed signature $\Sigma$ and a list of surface-level definitions $\vec{S}$ and outputs a new
well-formed signature $\Sigma'$.

We define $\Elab{\Sigma}{\vec{S}}{\Sigma'}$ in terms of three auxiliary judgments for elaborating
surface-language types and terms; in the bidirectional style, we divide term elaboration into a
checking judgment $\ChkTm{\Sigma;\Gamma}{\Prg{e}}{A}{\Sigma',M}$ taking a core type as input, and a
synthesis judgment $\SynTm{\Sigma;\Gamma}{\Prg{e}}{A}{\Sigma',M}$ producing a core type as output.
All three judgments take as input a signature $\Sigma$ and a context (telescope) over $\Sigma$, and
output a new signature along with a core type or term.

We represent a surface-level definition $S$ as a tuple:
\[
  \prn{\Kwd{def}~\vartheta:\Prg{A},\textit{abbrv?},\textit{abstr?},\brk{\kappa_1,\ldots\kappa_n},\Prg{e}}
\]
In this expression, $\vartheta$ is the name of the definiendum, $\Prg{A}$ is the surface-level type
of the definition, \textit{abbrv?} and \textit{abstr?} are flags governing whether $\vartheta$ is an
$\Kwd{abbreviation}$ (resp., is $\Kwd{abstract}$), $\brk{\kappa_1,\ldots,\kappa_n}$ are the names of
the definitions that $\vartheta$ unfolds, and $\Prg{e}$ is the surface-level definiens.

The elaboration judgment elaborates each surface definition in sequence:
\begin{mathpar}
  \inferrule{
    \Elab{\Sigma}{\vec{S}}{\Sigma_1}
    \\\\
    \ChkTp{\Sigma_1;\EmpCx}{\Prg{A}}{\Sigma_2,A}
    \\\\
    \ChkTm{\Sigma_2; \Conj{i\leq n}{\UnfTok{\kappa_i}}}{\Prg{e}}{A}{\Sigma_3,M}
    \\\\
    \Kwd{let}~{p}\coloneqq \Kwd{if}~\textit{abstr?}~\Kwd{then}~\textit{gensym}\,\prn{}~\Kwd{else}~\UnfTok{\vartheta}
    \\\\
    \Kwd{let}~{\boxslash}\coloneqq \Kwd{if}~\textit{abbrv?}~\Kwd{then}~\prn{=}~\Kwd{else}~\prn{\leq}
    \\\\
    \Kwd{let}~\Sigma_4 \coloneqq
    \Sigma_3,
    \Kwd{prop}\,p\mathrel{\boxslash} \Conj{i\leq n}{\UnfTok{\kappa_i}},
    \Kwd{const}\, \vartheta : \Ext{A}{p}{M}
  }{
    \Elab{\Sigma}{
      \vec{S},
      \prn{\Kwd{def}~\vartheta : \Prg{A}, \textit{abbrv?}, \textit{abstr?}, \brk{\kappa_1,\ldots,\kappa_n}, \Prg{e}}
    }{\Sigma_4}
  }
\end{mathpar}

\begin{remark}
  When a definition is marked $\Kwd{abstract}$, the name of the
  unfolding proposition is generated fresh so that it cannot be accessed by any future
  $\Kwd{unfold}$ declaration. Conversely, when a definition is marked as an
  $\Kwd{abbreviation}$, its unfolding proposition is defined to be equivalent to the
  conjunction of its dependencies rather than merely entailing its dependencies.
\end{remark}

The rules for term and type elaboration are largely standard: for instance, we
elaborate a surface dependent product to a core dependent product by
recursively elaborating the first and second components. We single out two
cases below: the boundary between checking and synthesis, and the
expression-level $\Kwd{unfold}$.
\begin{mathpar}
  \inferrule{
    \SynTm{\Sigma;\Gamma}{\Prg{e}}{A}{\Sigma_1;M}
    \\\\
    \Sigma_1;\Gamma\vdash \Con{conv}\,A\,B
  }{
    \ChkTm{\Sigma;\Gamma}{\Prg{e}}{B}{\Sigma_1;M}
  }
  \and
  \inferrule{
    \ChkTm{\Sigma;\Gamma,\UnfTok{\vartheta}}{\Prg{e}}{A}{\Sigma_1;M}
    \\
    \Kwd{let}~\chi\coloneqq \textit{gensym}\,\prn{}
    \\\\
    \Kwd{let}~\Sigma_2 \coloneqq
    \Sigma_1,
    \Kwd{const}\,\chi:\Prod{\Gamma}{\Ext{A}{\UnfTok{\vartheta}}{M}}
  }{
    \ChkTm{\Sigma;\Gamma}{
      \Kwd{unfold}~\vartheta~\Kwd{in}~\Prg{e}
    }{A}{
      \Sigma_2;
      \Out\Sub{\UnfTok{\vartheta}}\,\chi\brk{\Gamma}
    }
  }
\end{mathpar}

The first rule states that a term synthesizing a type $A$ can be checked against a type $B$ provided
that $A$ and $B$ are definitionally equal; in order to implement this rule algorithmically, we need
definitional equality to be decidable. Additionally, our (omitted) type-directed elaboration rules
are only well-defined if type constructors are injective up to definitional equality, \eg{}, $A \to
B = C \to D$ if and only if $A = C$ and $B = D$. %

Elaborating expression-level unfolding requires the ability to hoist a type to the top level by
iterating dependent products over its context, an operation notated $\Prod{\Gamma}$ above.
Because $\Gamma$ can hypothesize (the truth of) propositions, this operation relies crucially on the
presence of dependent products $\brc{p}\,A$.

\section{Case study: an implementation in \texorpdfstring{\cooltt{}}{cooltt}}
\label{sec:cooltt}

We have implemented our approach to controlled unfolding in the experimental \cooltt{} proof
assistant~\citep{cooltt}; \cooltt{} is an implementation of \emph{cartesian cubical type
theory}~\citep{angiuli:2021}, a computational version of homotopy type theory whose syntactic
metatheory is particularly well understood~\citep{huber:2019,sterling:phd,sterling:2021}. The existing support
for partial elements and extension types made \cooltt{} particularly hospitable for
experimentation with elaborating controlled unfolding to extension types. The following example
illustrates the use of controlled unfolding in \cooltt{}, where $\Con{path}~A~x~y$ is the cubical
notion of propositional equality ($x \equiv y$):

\iblock{
  \mhang{\Kwd{def}~{+} : \mathbb{N}\to\mathbb{N}\to\mathbb{N} \coloneqq}{
    \mrow{\Kwd{elim}}
    \mrow{\mid \Con{zero} \Rightarrow {n \Rightarrow n}}
    \mrow{\mid \Con{suc}~\brc{\_ \Rightarrow \mathit{ih}} \Rightarrow {n \Rightarrow \Con{suc}~\brc{\mathit{ih}\,n}}}
  }
  \row
  \mrow{\Kwd{unfold}~{+}}
  \mhang{\Kwd{def}~\Con{+0L}~ \prn{x : \mathbb{N}} : \Con{path}~\mathbb{N}~\brc{{+}~0~x}~x \coloneqq}{
    \mrow{i \Rightarrow x}
  }
  \row
  \mhang{
    {\Kwd{def}~\Con{+0R} : \prn{x : \mathbb{N}}\to \Con{path}~\mathbb{N}~\brc{{+}~x~0}~x \coloneqq}
  }{
    \mrow{\Kwd{elim}}
    \mrow{\mid \Con{zero} \Rightarrow \Con{+0L}~0}
    \mhang{\mid \Con{suc}~\brc{x\Rightarrow \mathit{ih}} \Rightarrow}{
      \mrow{\Kwd{equation}~\mathbb{N}}
      \mrow{
        \begin{array}[t]{ll}
          \mid {+}~0~\brc{\Con{suc}~y} & {=}\brk{\Con{+0L}~\brc{\Con{suc}~y}}\\
          \mid \Con{suc}~\brc{{+}~x~0} & {=}\brk{i \Rightarrow \Con{suc}~\brc{\mathit{ih}~i}}\\
          \mid \Con{suc}~x~
        \end{array}
      }
    }
  }
}

This example follows a common pattern: we prove basic computational laws ($\Con{+0L}$) by unfolding
a definition, and then in subsequent results ($\Con{+0R}$) use these lemmas abstractly rather than
unfolding. Doing so controls the size and readability of proof goals, and explicitly demarcates
which parts of the library depend on the definitional behavior of a given function.

We have also implemented the derived forms for expression-level unfolding:

\iblock{
  \mrow{\Kwd{def}~\Con{two} : \mathbb{N} \coloneqq {+}~1~1}
  \mrow{
    \Kwd{def}~\Con{thm} : \Con{path}~\mathbb{N}~\Con{two}~2\coloneqq
    \Kwd{unfold}~\Con{two}~{+}~\Kwd{in}~i \Rightarrow 2
  }
  \mhang{
    \Kwd{def}~\Con{thm\mhyphen{}is\mhyphen{}refl} :
    \Con{path\text{-}p}~\brc{i\Rightarrow\Con{path}~\mathbb{N}~\Con{two}~\brc{\Con{thm}~i}}~\brc{i\Rightarrow \Con{two}}~\Con{thm} \coloneqq
  }{
    \mrow{i~j\Rightarrow \Kwd{unfold}~\Con{two}~{+}~\Kwd{in}~2}
  }
  \mhang{
    \Kwd{def}~\Con{thm\mhyphen{}is\mhyphen{}refl'} :
    \Con{path}~\brc{\Con{path}~\mathbb{N}~\Con{two}~2}~\brc{i\Rightarrow\Kwd{unfold}~\Con{two}~{+}~\Kwd{in}~\Con{two}}~\Con{thm}\coloneqq
  }{
    \mrow{i~j\Rightarrow \Kwd{unfold}~\Con{two}~{+}~\Kwd{in}~2}
  }
}

The third and fourth declarations above illustrate two strategies in \cooltt{} for dealing with a
dependent type whose well-formedness depends on an unfolding; in $\Con{thm\mhyphen{}is\mhyphen{}refl}$ we use a
dependent path type but only unfold in the definiens, whereas in $\Con{thm\mhyphen{}is\mhyphen{}refl'}$ we use a
non-dependent path type but must unfold in both the definiens and in its type.

Our \cooltt{} implementation deviates in a few respects from the presentation in this
paper: in particular, the propositions $\UnfTok{\kappa}$ are
represented by abstract elements $i_\kappa : \mathbb{I}$ of the interval via
the embedding $\mathbb{I}\hookrightarrow\mathbb{F}$ sending $i$ to $\prn{i=\Sub{\mathbb{I}}1}$.

\cooltt{} utilizes a standalone library to compute entailment of cofibrations called
Kado~\citep{kado}, created by Kuen-Bang Hou (Favonia). To support our experiment, Favonia modified
Kado to support inequalities of dimension variables $i\leq\Sub{\mathbb{I}} j$ in addition to the
cofibrations needed for \cooltt{}'s core theory. As a result, the modifications to \cooltt{} were
quite modest. After the changes to Kado---which could in principle be reused in any proof assistant
for the same purpose---the entire change required only a net increase of 996 lines of OCaml
code. %
\section{The metatheory of \texorpdfstring{\TTProp{}}{TT\_P}}
\label{sec:normalization}

In Section~\ref{sec:elaboration} we described an algorithm elaborating a surface language with controlled
unfolding to \TTProp{}. In order to actually execute our algorithm, it is necessary to decide the
definitional equality of types in \TTProp{}; as is often the case in type theory, type dependency
ensures that deciding equality for types also requires us to decide the equality of terms. In order
to implement our elaboration algorithm, we therefore prove a \emph{normalization} result for
\TTProp{}.

At its heart, a normalization algorithm is a computable bijection between equivalence classes of
terms up to definitional equality and a collection of normal forms. By ensuring that the equality of
normal forms is evidently decidable, this yields an effective decision procedure for definitional
equality. In our case, we attack normalization through a \emph{synthetic} and \emph{semantic}
approach to normalization by evaluation called synthetic Tait computability
\citep{sterling:modules:2021,sterling:2021,sterling:phd,sterling:2025:grothendieck} or STC.

\paragraph*{Neutral forms for \TTProp{}}
The semantic analysis of normalization by evaluation rests on the observation of \citet{fiore:2002} that normal forms, though not stable under arbitrary substitutions, are nonetheless stable under \emph{renamings} --- substitutions that replace variables with variables (not necessarily injective). Therefore, decisive aspects of the normalization algorithm can be expressed \emph{internally} to a topos of variable sets (presheaves) over the category of contexts and renamings; in order to instrument the semantic normalization algorithm with its proof of correctness, one passes to a larger topos obtained from the former by gluing. Synthetic Tait computability then instantiates the standard topos model of Martin-L\"of type theory to substantially simplify various details that would otherwise be exceedingly tedious by means of a form of higher-order abstract syntax.

This appealingly simple story for normalization is substantially complicated by the boundary law
for extension types:
\[
  \inferrule{
    \IsTrue{p}
    \\
    \IsTm[\RCx{\Gamma}{p}]{a\Sub{p}}{A}
    \\
    \IsTm{a}{\Ext{A}{p}{a\Sub{p}}}
  }{
    \EqTm{\Out_p\,a}{a\Sub{p}}{A}
  }
\]

When defining normal forms for \TTProp{}, we might naively add a neutral form $\NeOut_p$ to represent
$\Out_p$. In order to ensure that normal and neutral forms correspond bijectively with equivalence
classes of terms, however, we should only allow $\NeOut_p$ to be applied in a context where $p$ is
not true; if $p$ were true, $\Out_p\,{a}$ is already represented by the normal form for $a\Sub{p}$.

A similar problem arises in the context of cubical type theory~\citep{cohen:2017,angiuli:2021} where
some equalities apply precisely when two \emph{dimensions} coincide. The same problem arises:
either renamings must exclude substitutions that identify two dimension terms, or neutral forms
will not be stable under renamings. 
In their proof of normalization for cubical type theory, \citet{sterling:2021} refined neutral forms to account for this tension by introducing
\emph{stabilized neutrals}. Rather than cutting down on renamings, they expand the class of
neutrals by allowing ``bad'' neutrals akin to $\NeOut_p\,e$ in a context where $p$ is
true. They then associate each neutral form with a \emph{frontier of instability}: a proposition
that becomes true when the neutral is no longer ``stuck''. Crucially, although well-behaved neutrals may not
be stable under renamings, the frontier of instability \emph{is} stable, and can therefore be incorporated into
the internal language.

We adapt Sterling and Angiuli's stabilized neutrals to the simplified setting of
\TTProp{} and establish its normalization theorem. In so doing, we
refine the approach of \opcit to obtain a fully constructive\footnote{By \emph{constructive}, we mean something that can be carried out in an elementary topos with a natural numbers object.} normalization
proof. We also carefully spell out the details of the universe in the
normalization model, correcting an oversight in an earlier revision of Sterling's
dissertation~\citep{sterling:phd}.

\subsection{Type theories as categories with representable maps}

While any number of logical frameworks are available (generalized algebraic
theories~\citep{cartmell:1978}, essentially algebraic theories~\citep{freyd:1972}, locally cartesian
closed categories~\citep{gratzer-sterling:lcccs:2020}, \etc{}), Uemura's \emph{categories with
representable maps}~\citep{uemura:2023,uemura:phd} are particularly attractive because they express
exactly the binding and dependency structure needed for type theory: a second-order version of
generalized algebraic theories.

\begin{definition}
  A \DefEmph{category with representable maps (CwR)} $\CC$ is a finitely complete category equipped with a
  pullback-stable class of \DefEmph{representable maps} $\mathcal{R} \subseteq \Arr\prn{\CC}$ such that
  pullback along $f \in \mathcal{R}$ has a right adjoint (dependent product along $f$).
\end{definition}

\begin{definition}
  A \DefEmph{morphism of CwRs} is a functor between the underlying categories that
  preserves finite limits, representability of maps, and dependent products
  along representable maps.
\end{definition}

\begin{definition}
  CwRs, morphisms between them, and natural isomorphisms assemble into
  a $(2,1)$-category $\CWR$.
\end{definition}

Uemura's logical framework axiomatizes the category of judgments of \TTProp{}
as a particular category with representable maps $\TT$. The finite limit
structure of $\TT$ encodes substitution as well as equality judgments, while
the class of representable maps carves out those judgments that may be
hypothesized. \citet{uemura:2023} develops a syntactic method for
presenting a CwR as a signature within a variant of extensional type theory,
which he has rephrased in terms of \emph{second order generalized algebraic
theories} in his doctoral dissertation~\citep{uemura:phd}. Although we will use
the type-theoretic presentation for convenience, the difference between these
two accounts is only superficial.

Each judgment of \TTProp{} is rendered as a (dependent) sort while operators are modeled by elements
of the given sorts. In order to record whether a given judgment may be hypothesized, the sorts of
the type theory are stratified by \emph{meta}-sorts $\star \subseteq \Box$ where $A : \star$
signifies that $A$ is a representable sort (\ie a context-former) and can be hypothesized, whereas
$B : \Box$ cannot parameterize a framework-level dependent product.

\begin{proposition}
  \label{prop:normalization:tt-universal-prop}
  Let $\mathbb{T}$ be the free category with representable maps generated by a given logical
  framework signature;
  then the groupoid of CwR functors $\Hom{\CWR}{\TT}{\EE}$ is equivalent to
  the groupoid of interpretations of the signature within $\EE$.
\end{proposition}

We will often refer to a category with representable maps $\TT$ as a
type theory; indeed, as the category of judgments of a given type theory,
$\TT$ is a suitable invariant replacement for it.

Proposition~\ref{prop:normalization:tt-universal-prop} describes the
universal property of a type theory generated by a given signature in a logical
framework. Type theories \emph{qua} CwRs thus give rise to a form of
functorial semantics in which algebras (interpretations) arrange into a
\emph{groupoid} of CwR functors $\Hom{\CWR}{\TT}{\EE}$.

This is an appropriate setting for studying the syntax of type theory, but it
is somewhat inappropriate for studying the \emph{semantics} of type theory---in
which one expects models to correspond to structured
CwFs~\citep{dybjer:1996} or natural models~\citep{awodey:2018}, which
themselves arrange into a (2,1)-category. The second notion of
functorial semantics, developed by Uemura in his doctoral
dissertation~\citep{uemura:phd}, is a generalization of the theory of CwFs and
pseudo-morphisms between them~\citep{clairambault:2014,newstead:2018}.

Note that we may always regard a presheaf category $\Psh{\CC}$ as a CwR with the representable maps
being representable natural transformations, \ie families of presheaves whose fibers at
representables are representable~\citep{awodey:2018}.

\begin{definition}\label{def:taichi-model}
  A \DefEmph{model} of a type theory $\TT$ is a category $\MMod_\diamond$ together with a CwR
  functor $\Mor[\MMod]{\TT}{\Psh{\MMod_\diamond}}$.
\end{definition}

Models are arranged into a (2,1)-category $\MOD{\TT}$ (see
Appendix~\ref{app:category-of-models}). Essentially, a morphism of models $\Mor{\MMod}{\NMod}$
is given by a functor $\Mor[\alpha_\diamond]{\MMod_\diamond}{\NMod_\diamond}$ together
with a natural transformation
$\Mor{\MMod}{\AtmFig^*\NMod}\in\Hom{\CWR}{\TT}{\Psh{\MMod_\diamond}}$ that
preserves context extensions up to isomorphism; an isomorphism between morphisms of models is a
natural isomorphism between the underlying functors satisfying an additional property.

For each CwR $\TT$, Uemura has shown the following theorem:
\begin{proposition}
  The (2,1)-category of models $\MOD{\TT}$ has a bi-initial object $\IMod$, whose category of contexts $\IMod_\diamond$ is the smallest full subcategory of $\TT$ closed under the terminal object and pullbacks along representable maps.
\end{proposition}

\begin{remark}
  If one takes $\TT$ to be \eg{}, Martin-L{\"o}f type theory, the bi-initial model $\IMod$ can be realized by the familiar initial CwF built from the category of contexts.
\end{remark}

\subsection{Encoding \texorpdfstring{\TTProp{}}{TT\_P} in the logical framework}
\label{sec:normalization:sig}

We begin by defining the signature for a category with representable maps
$\TT_0$ containing exactly the bare judgmental structure of \TTProp, namely the
propositions and the judgments for types and terms. In our signature, we make
liberal use of the Agda-style notation for implicit arguments. As always, $p$
ranges over $\PP$.

\iblock{
  \mrow{\gl{p}:\star}
  \mrow{\Ty : \square}
  \mrow{\Tm : \Ty\TTo \star}
  \mrow{\_:\brc{u,v:\gl{p}}\TTo u = v}
  \mrow{\_:\brc{\_ : \gl{\Conj{i<n}p_i}}\TTo \gl{p_k}}
  \mrow{\_:\brc{\_:\gl{p_i},\dots}\TTo\gl{\Conj{i<n} p_i}}
}

Note that already this signature encodes the necessary theory of propositions for $\TTProp$. For
instance, if $p \le q$ in $\PP$ then a combination of the final two implications in the signature
implies $\gl{p} \to \gl{q}$. We next extend the above to include the type formers of $\TTProp$,
writing $\TT$ for the CwR generated by the full signature.

\begin{notation}\label{notation:lf:implicit-pi}
  Given $X : \brc{\_ : \gl{p}}\TTo \square$, we will write $\brc{p}\, X$ to further abbreviate the
  Agda-style implicit function space $\brc{\_ : \gl{p}}\to X$. Note that $\brc{p}\,X$ still
  associates to the right and so $\brc{p} A \to B$ signifies $\brc{p} \prn{A \to B}$.
\end{notation}

For instance, the following constants specify the rules of extension types given in
Section~\ref{sec:elaboration:core-calculus}:

\iblock{
  \mrow{\ExtConst_p : \prn{A : \Ty}\,\prn{a : \brc{p}\, \Tm\,{A}}\TTo \Ty}
  \mrow{\Con{in}_p : \prn{A : \Ty}\,\prn{a : \brc{p}\, \Tm\,{A}}\,\prn{u : \Tm\,{A}}\,\brc{\_ : \brc{p}\, u=a} \TTo \Tm\,\prn{\ExtConst_p\,A\,a}}
  \mrow{\Con{out}_p : \prn{A:\Ty}\,\prn{a:\brc{p}\,\Tm\,{A}}\,\prn{u:\Tm\,\prn{\ExtConst_p\,A\,a}}\TTo \Tm\,{A}}
  \mrow{\_ : \prn{A:\Ty}\,\prn{a:\brc{p}\,\Tm\,{A}}\,\prn{u:\Tm\,\prn{\ExtConst_p\,A\,a}}\,\brc{\_:\gl{p}} \TTo \Con{out}_p\,A\,a\,u = a}
  \mrow{\_ : \prn{A:\Ty}\,\prn{a:\brc{p}\,\Tm\,{A}}\,\prn{u:\Tm\,{A}}\,\brc{\_:\brc{p}\, u = a}\TTo \Con{out}_p\,A\,a\,\prn{\Con{in}_p\,A\,a\,u} = u}
  \mrow{\_ : \prn{A:\Ty}\,\prn{a:\brc{p}\,\Tm\,{A}}\,\prn{u:\Tm\,\prn{\ExtConst_p\,A\,a}}\TTo \Con{in}_p\,A\,a\,\prn{\Con{out}_p\,A\,a\,u} = u}
}

The full list of non-standard constants is specified in Figure~\ref{fig:full-lf}. Once the signature is
complete, we obtain from Uemura's framework a category with representable maps $\TT$ together with a
bi-initial model $\IMod$.

\begin{figure*}
\iblock{
  \mrow{\Ty : \square}
  \mrow{\Tm : \Ty\TTo \star}
  \row
  \mrow{\gl{p}:\star}
  \mrow{\_:\brc{u,v:\gl{p}}\TTo u = v}
  \mrow{\_:\brc{\_ : \gl{\Conj{i<n}p_i}}\TTo \gl{p_k}}
  \mrow{\_:\brc{\_:\gl{p_i},\dots}\TTo\gl{\Conj{i<n} p_i}}

  \row

  \mrow{\ExtConst_p : \prn{A : \Ty}\,\prn{a : \brc{p}\, \Tm\,{A}}\TTo \Ty}
  \mrow{\Con{in}_p : \prn{A : \Ty}\,\prn{a : \brc{p}\, \Tm\,{A}}\,\prn{u : \Tm\,{A}}\,\brc{\_ : \brc{p}\, u=a} \TTo \Tm\,\prn{\ExtConst_p\,A\,a}}
  \mrow{\Con{out}_p : \prn{A:\Ty}\,\prn{a:\brc{p}\,\Tm\,{A}}\,\prn{u:\Tm\,\prn{\ExtConst_p\,A\,a}}\TTo \Tm\,{A}}
  \mrow{\_ : \prn{A:\Ty}\,\prn{a:\brc{p}\,\Tm\,{A}}\,\prn{u:\Tm\,\prn{\ExtConst_p\,A\,a}}\,\brc{\_:\gl{p}} \TTo \Con{out}_p\,A\,a\,u = a}
  \mrow{\_ : \prn{A:\Ty}\,\prn{a:\brc{p}\,\Tm\,{A}}\,\prn{u:\Tm\,{A}}\,\brc{\_:\brc{p}\, u = a}\TTo \Con{out}_p\,A\,a\,\prn{\Con{in}_p\,A\,a\,u} = u}
  \mrow{\_ : \prn{A:\Ty}\,\prn{a:\brc{p}\,\Tm\,{A}}\,\prn{u:\Tm\,\prn{\ExtConst_p\,A\,a}}\TTo \Con{in}_p\,A\,a\,\prn{\Con{out}_p\,A\,a\,u} = u}

  \row

  \mrow{\PartialConst_p : \prn{A : \brc{p}\,\Ty} \TTo \Ty}
  \mrow{\Con{lam}_p : \prn{A : \Ty}\,\prn{a : \brc{p}\, \Tm\,{A}} \TTo \Tm\,\prn{\PartialConst_p\,A}}
  \mrow{\Con{app}_p : \prn{A:\Ty}\,\prn{u:\Tm\,\prn{\PartialConst_p\,A}} \TTo \brc{p}\,\Tm\,{A}}
  \mrow{\_ : \prn{A:\Ty}\,\prn{a:\brc{p}\,\Tm\,{A}} \TTo \Con{app}_p\,A\,\prn{\Con{lam}_p\,A\,a} = a}
  \mrow{\_ : \prn{A:\Ty}\,\prn{a:\Tm\,\prn{\PartialConst_p\,A}} \TTo \Con{lam}_p\,A\,\prn{\Con{app}_p\,A\,a} = a}
}
  \caption{The non-standard aspects of the LF signature for \TTProp.}
  \label{fig:full-lf}
\end{figure*}

\subsection{The atomic figure shape and its universal property}
\label{sec:normalization:renamings}

For each context $\Gamma$ and type $\IsTy[\Gamma]{A}$, it is possible to
axiomatize the normal forms of type $A$; unfortunately, this assignment of sets
of normal forms does not immediately extend to a presheaf on the category of
contexts $\IMod_\diamond$, precisely because normal forms are not \emph{a
priori} closed under substitution! In fact, closing normal forms under
substitution is the purpose of normalization, so we are not
able to assume it beforehand.

Normal forms \emph{are}, however, closed under substitutions of variables for
variables (often called \emph{structural renamings}), and in our case we shall
be able to close them additionally under the ``phase transitions''
$\Mor{\Gamma,\gl{p}}{\Gamma,\gl{q}}$ when $\IsTrue[\Gamma,p]{q}$ is derivable.
We shall refer to these substitutions as \DefEmph{atomic
substitutions}, and we wish to organize them into a category.

It is possible to inductively define a category of ``atomic contexts'' whose objects are those of
$\IMod_\diamond$ and whose morphisms are atomic substitutions, but this construction obscures a
beautiful and simple (2,1)-categorical universal property first exposed by \citet{bocquet-kaposi-sattler:2021} that leads to a more modular proof. To explicate this
universal property, first note that the theory $\TT_0$ axiomatizes exactly the structure of
variables and phase transitions, and that the initial model $\IMod$ of $\TT$ is, by restriction
along $\Mor{\TT_0}{\TT}$, also a model of $\TT_0$.

\begin{definition}
  \label{def:normalization:renamings}
  An \DefEmph{atomic substitution model} over a fixed $\TT$-model $\MMod$ is given by a model $\AtmMod$ of the
  bare judgmental theory $\TT_0$, together with a morphism of models
  $\Mor[\alpha]{\AtmMod}{\MMod}$ in $\MOD{\TT_0}$ such that
  $\Mor[\alpha\Sub{\Ty}]{\AtmMod\prn{\Ty}}{\AtmFig^*\prn{\MMod\prn{\Ty}}}\in
  \Psh{\AtmMod_\diamond}$ is an isomorphism.
\end{definition}

Atomic substitution model over $\MMod$ arrange themselves into a (2,1)-category, a full subcategory of $\MOD{\TT_0}\downarrow\MMod$. The following result is due to \citet{bocquet-kaposi-sattler:2021}.

\begin{proposition}\label{prop:biinitial-subst-model}
  The bi-initial atomic substitution model $\prn{\AtmMod, \Mor[\alpha]{\AtmMod}{\IMod}}$ over $\IMod$ exists.
\end{proposition}

When $\prn{\AtmMod, \Mor[\alpha]{\AtmMod}{\IMod}}$ is the bi-initial atomic substitution model over $\IMod$ as in Proposition~\ref{prop:biinitial-subst-model}, we shall refer to an object $\Gamma\in\AtmMod$ as an \emph{atomic context} and a morphism $\Mor[\gamma]{\Delta}{\Gamma}$ in $\AtmMod$ as an \emph{atomic substitution}. We shall assume without loss of generality that $\AtmMod\prn{\Ty} = \AtmFig^*\IMod\prn{\Ty}$ so that the component $\alpha_{\Ty}$ is the identity map.

\subsection{Computability spaces by gluing along the atomic figure shape}

We shall use the bi-initial atomic substitution model over $\IMod$ as a \DefEmph{figure
shape} in the sense of \citet[\S4.3]{sterling:phd} to instantiate
synthetic Tait computability. Here we transition into the 2-category of
Grothendieck topoi, geometric morphisms, and geometric transformations, guided
by a \DefEmph{phase distinction} between ``object-space'' and
``meta-space''~\citep{sterling:phd};\footnote{In his doctoral dissertation, Sterling referred to ``object-space'' and ``meta-space'' as \emph{syntactic} and \emph{semantic} respectively~\citep{sterling:phd}. However, there are compelling reasons to consider object-space more semantic than meta-space (in which various admissibilities hold that will not be preserved by homomorphisms of models), so we have changed terminology to avoid confusion.} object-space refers to the object
language embodied in the model $\IMod$, whereas meta-space refers to the
metalanguage embodied in the model $\AtmMod$. Later on, we will construct a
\emph{glued} topos in which we may speak of constructs that have extent in both
object-space and meta-space. We follow \citet{anel:2021} and
\citet{vickers:2007} in emphasizing the distinction between a topos
$\XTop$ and the category of sheaves $\Sh{\XTop}$ presenting it:

\begin{definition}
  We denote by $\ObjTop$ and $\MetaTop$ the \DefEmph{object-space} and \DefEmph{meta-space topoi}
  respectively, with underlying categories of sheaves $\Sh{\ObjTop} = \Psh{\IMod_\diamond}$ and
  $\Sh{\MetaTop} = \Psh{\AtmMod_\diamond}$.
\end{definition}

\begin{definition}
  The functor $\Mor[\alpha_\diamond]{\AtmMod_\diamond}{\IMod_\diamond}$
  gives rise under precomposition to a continuous and cocontinuous
  functor $\Mor{\Psh{\IMod_\diamond}}{\Psh{\AtmMod_\diamond}}$, that shall
  serve as the inverse image part of an (essential) geometric morphism
  $\Mor[\AtmFig]{\MetaTop}{\ObjTop}$ named the \DefEmph{atomic figure shape}.
\end{definition}

That $\Mor[\AtmFig]{\MetaTop}{\ObjTop}$ is \emph{essential} means that its inverse image $\Mor[\AtmFig^*]{\Sh{\ObjTop}}{\Sh{\MetaTop}}$ has a left adjoint $\Mor[\AtmFig_!]{\Sh{\MetaTop}}{\Sh{\ObjTop}}$; from the point of view of presheaves, this is precisely the Yoneda extension of $\Mor[\alpha_\diamond]{\AtmMod_\diamond}{\IMod_\diamond}$ as depicted below:
\[
  \DiagramSquare{
    nw = \AtmMod_\diamond,
    ne = \IMod_\diamond,
    sw = \Psh{\AtmMod_\diamond},
    se = \Psh{\IMod_\diamond},
    west = \Yo{\AtmMod_\diamond},
    east = \Yo{\IMod_\diamond},
    north = \alpha_\diamond,
    south = \AtmFig_!,
    west/style = embedding,
    east/style = embedding,
    south/style = {exists,->},
  }
\]

\begin{definition}\label{def:gluing-topos}
  We denote by $\GlTop$ the \emph{closed mapping
  cylinder}~\citep{johnstone:topos:1977} of the geometric morphism
  $\Mor[\AtmFig]{\MetaTop}{\ObjTop}$; in other words, $\Sh{\GlTop}$ is the comma
  category $\COMMA{\Sh{\MetaTop}}{\AtmFig^*}$. We will write
  $\Mor|open immersion|[\OpenIncl]{\ObjTop}{\GlTop}$ and $\Mor|closed
  immersion|[\ClIncl]{\MetaTop}{\GlTop}$ for the open and closed
  subtopos immersions.
\end{definition}

Following \citet{sterling:2025:grothendieck}, we shall refer to a sheaf on $\GlTop$ as a \DefEmph{computability space}. A computability space $X\in\Sh{\GlTop}$ is then identified with a family $\Mor[\pi_X]{\ClIncl^*X}{\AtmFig^*\OpenIncl^*X}$ in $\Sh{\MetaTop}$. Because the assignment  $\pi_X$ is natural in computability spaces $X$, it corresponds to a 2-cell $\Mor[\pi]{\OpenIncl\circ\AtmFig}{\ClIncl}$ in the 2-category of Grothendieck topoi. The universal property of $\GlTop$ is then expressed by the fact that $\Mor[\pi]{\OpenIncl\circ\AtmFig}{\ClIncl}$ is a \emph{co-comma} cell in the 2-category of Grothendieck topoi:
\[
  \begin{tikzpicture}[diagram]
    \SpliceDiagramSquare{
      nw = \MetaTop,
      sw = \MetaTop,
      ne = \ObjTop,
      se = \GlTop,
      east = \OpenIncl,
      south = \ClIncl,
      north = \AtmFig,
      east/style = open immersion,
      south/style = closed immersion,
      west/style = double,
    }
    \path (sw) to node {${\Uparrow}\pi$} (ne);
  \end{tikzpicture}
\]

\begin{remark}[Relation to Kripke computability predicates]
  Unraveling Definition~\ref{def:gluing-topos}, a computability space is precisely a \emph{family} $X'$ of presheaves on $\AtmMod_\diamond$ indexed in the restriction of a given presheaf $X$ on $\IMod_\diamond$ along $\Mor[\alpha_\diamond]{\AtmMod_\diamond}{\IMod_\diamond}$. When the family $X'$ is valued in \emph{subterminal} presheaves and the base $X$ is representable, we have precisely the classical notion of a Kripke computability predicate~\citep{jung-tiuryn:1993}; a \emph{computability space} in our sense is then a generalised, proof-relevant version of a Kripke computability predicate.
\end{remark}

\subsubsection{Reflection of object and meta-space}\label{sec:normalization:reflection:external}
By definition, the inverse image functors $\OpenIncl^*,\ClIncl^*$ have fully faithful right adjoints $\EmbMor[\OpenIncl_*]{\Sh{\ObjTop}}{\Sh{\GlTop}}$ and $\EmbMor[\ClIncl_*]{\Sh{\MetaTop}}{\Sh{\GlTop}}$ respectively. These are computed as follows:
\begin{align*}
  \OpenIncl_*E &= \prn{E, \Mor[1_{\AtmFig^*E}]{\AtmFig^*E}{\AtmFig^*E}}\\
  \ClIncl_*A &= \prn{\ObjTerm{\Sh{\ObjTop}}, \Mor[!_{A}]{A}{\ObjTerm{\Sh{\MetaTop}}\cong \AtmFig^*\ObjTerm{\Sh{\ObjTop}}}}
\end{align*}

Thus the adjunctions $\OpenIncl^*\dashv\OpenIncl_*$ and $\ClIncl^*\dashv\ClIncl_*$ exhibit $\Sh{\ObjTop}$ and $\Sh{\MetaTop}$ as \emph{reflective} subcategories of $\Sh{\GlTop}$.

\begin{enumerate}
  \item The essential image of the reflective embedding $\EmbMor[\OpenIncl_*]{\Sh{\ObjTop}}{\Sh{\GlTop}}$ is spanned by computability spaces $X$ for which $\Mor[\pi_X]{\ClIncl^*X}{\AtmFig^*\OpenIncl^*X}$ is an isomorphism, \ie so that $\pi_X$ is terminal in the slice $\Sh{\MetaTop}\downarrow\AtmFig^*\OpenIncl^*X$.
  \item The essential of image of the reflective embedding $\EmbMor[\ClIncl_*]{\Sh{\MetaTop}}{\Sh{\GlTop}}$ is spanned by computability spaces $X$ such that $\OpenIncl^*X$ is terminal in $\Sh{\ObjTop}$.
\end{enumerate}

\begin{definition}[Vocabulary for reflective subcategories]
  When a computability space lies in the essential image of $\EmbMor[\OpenIncl_*]{\Sh{\ObjTop}}{\Sh{\GlTop}}$, we shall refer to it as \DefEmph{lying in object-space}. Likewise, when a computabiltiy space lies in the essential image of $\EmbMor[\ClIncl_*]{\Sh{\MetaTop}}{\Sh{\GlTop}}$, we shall say that it \DefEmph{lies in meta-space}.
\end{definition}

\subsubsection{Coreflection of object and meta-space}\label{sec:normalization:coreflection:external}

Both the open and closed immersions are \emph{essential} morphisms of topoi, in the sense that we have additional (necessarily fully faithful) left adjoints $\OpenIncl_!\dashv\OpenIncl^*\colon\Sh{\GlTop}\to \Sh{\ObjTop}$ and $\ClIncl_!\dashv \ClIncl^*\colon\Sh{\GlTop}\to\Sh{\MetaTop}$ that are computed as follows:
\begin{align*}
  \OpenIncl_!E &= \prn{E, \Mor[!_{\AtmFig^*E}]{\ObjInit{\Sh{\MetaTop}}}{\AtmFig^*E}}
  \\
  \ClIncl_!A &= \prn{\AtmFig_!A, \Mor[\eta_A]{A}{\AtmFig^*\AtmFig_!A}}
\end{align*}

Thus $\Sh{\ObjTop}$ and $\Sh{\MetaTop}$ are not only reflective in
$\Sh{\GlTop}$ --- they are also \emph{coreflective}.

\subsection{The language of synthetic Tait computability}
\label{sec:normalization:stc}

As $\ObjTop$ and $\MetaTop$ are both subtopoi of $\GlTop$, their reflections (Section~\ref{sec:normalization:reflection:external}) can be expressed
in the internal language of $\Sh{\GlTop}$ by means of a pair of complementary
lex idempotent monads ($\Open$,$\Closed$). The internal language of
$\Sh{\ObjTop}$ is presented by the $\Open$-modal or \emph{object-space} types
and $\Sh{\MetaTop}$ is presented by $\Closed$-modal or \emph{meta-space} types.
Because they form an open/closed partition, these modal subuniverses admit a
particularly simple formulation:

\begin{therm}
  There exists a proposition $\Syn : \Prop$ such that
  \begin{enumerate}
  \item a type $X$ is $\Open$-modal / object-space iff $X \to \prn{\Syn \to X}$ is an isomorphism;
  \item a type $X$ is $\Closed$-modal / meta-space iff $\Syn\times X\to \Syn$ is an isomorphism.
  \end{enumerate}
\end{therm}

\begin{remark}
In fact, the coreflection of $\Sh{\ObjTop}$ in $\Sh{\GlTop}$ lifts smoothly into the internal language (though we shall not use this fact) by the idempotent comonadic modality $\square\dashv\Open$ that sends ${X}$ to the product $\square{X} = \Syn\times X$. On the other hand, the coreflection of $\Sh{\MetaTop}$ in $\Sh{\GlTop}$ cannot be expressed directly in the internal language.
\end{remark}

\begin{notation}
  We will use extension types $\Ext{A}{\phi}{a}$ in the internal language of $\Sh{\GlTop}$ as
  realized by the subset comprehension of topos logic, treating their introduction and elimination
  rules silently. Here $\phi$ will be an element of the subobject classifier, in contrast to the
  situation in our object language, where it ranged over fixed proposition symbols.
\end{notation}

\begin{remark}
  We assume a subuniverse $\DecProp\subseteq\Prop$ of the subobject classifier that is closed under
  finite disjunctions and contains $\Syn$; then $\DecProp$ will ultimately be a subuniverse spanned by
  pointwise/externally decidable propositions~\citep{angiuli:2021}, but this fact will not play a
  role in the synthetic development.
\end{remark}

\begin{notation}
  We will reuse Notation~\ref{notation:lf:implicit-pi} and write $\brc{\Syn}\,A$ rather than
  $\brc{\_ : \Syn} \to A$ when $A : \brc{\_:\Syn}\to\Uni$.
\end{notation}

As a presheaf topos, $\Sh{\GlTop}$ inherits a hierarchy of cumulative universes
$\Uni[i]$, each of which supports the \DefEmph{strict gluing} or
\DefEmph{(mixed-phase) refinement}
type~\citep{gratzer:universes:2022}: a version of the
dependent sum of a family of meta-space types indexed in an object-space type
$A$ that \emph{additionally} restricts within
object-space to exactly $A$:
\[
  \mprset{center}
  \inferrule{
    A : \brc{\Syn}\, \Uni[i]
    \\
    B : \prn{\brc{\Syn}\, A} \to \Uni[i]
    \\
    \brc{\Syn} \prn{a : A} \to \prn{B\,a \cong \ObjTerm{}}
  }{
    \Glue{x}{A}{B\,x} : \Ext{\Uni[i]}{\Syn}{A}
    \\
    \Con{gl} : \Ext{\prn{\prn{x : \brc{\Syn}\, A}\times B\,x} \cong \Glue{x}{A}{B\,x}}{\Syn}{\Proj{1}}
  }
\]

\begin{remark}
  \label{rem:normalization:isos}
  In topos logic, it is a \emph{property} for a function to have an inverse; thus we have
  conveniently packaged the introduction and elimination rules for $\Glue{x}{A}{B\,x}$ into a single
  function $\Con{gl}$ that is assumed to be an isomorphism.
\end{remark}

\begin{notation}
  \label{not:normalization:mkglue}
  We write $\MkGlue{a}{b}$ for $\Con{gl}\,\prn{a,b}$ and $\Unglue\,x$ for
  $\Proj{2}\prn{\Con{gl}\Inv x}$. When constructing particularly complex inhabitants of
  $\Glue{x}{A}{B\,x}$ we will avail ourselves of copattern matching notation and write the following
  instead of $c = \MkGlue{a}{b}$:
  \iblock{
    \mrow{\Syn \hookrightarrow c = a}
    \mrow{\Unglue\,c = b}
  }
\end{notation}

Both $\Open$ and $\Closed$ induce reflective subuniverses
$\EmbMor{\Uni_{\Open}^i,\Uni_{\Closed}^i}{\Uni[i]}$ spanned by modal types, and these universes are
themselves modal. Following \citet{sterling:phd}, we use strict gluing to choose these
universes with additional strict properties:
\[
  \Uni_{\Open}^i : \Ext{\Uni[i+1]}{\Syn}{\Uni[i]}
  \qquad
  \Uni_{\Closed}^i : \Ext{\Uni[i+1]}{\Syn}{\ObjTerm{}}
\]
Furthermore, the inclusion $\EmbMor{\Uni_{\Open}^i}{\Uni[i]}$ restricts to the identity under
$\Syn$. With the modal universes to hand, we may choose $\Open : \Uni[i] \to \Uni[i]$ and
$\Closed : \Uni[i] \to \Uni[i]$ to factor through $\Uni_\Open^i$ and $\Uni_\Closed^i$
respectively. Henceforth we will suppress the inclusions
$\EmbMor{\Uni_{\Open}^i,\Uni_{\Closed}^i}{\Uni[i]}$ and write \eg{}
$\Open : \Uni[i] \to \Uni_\Open^i$ for the reflections.

\begin{remark}
  The strict gluing types, modal universes, and their modal reflections can be
  chosen to commute strictly with the liftings $\Mor{\Uni[i]}{\Uni[i + 1]}$.
\end{remark}

The interpretation of the \TTProp{} signature within $\Sh{\ObjTop}$
internalizes into $\Sh{\GlTop}$ as a sequence of constants valued in the
subuniverse $\Uni_\Open^0$; for instance, we have:

\iblock{
  \mrow{
    \Ty : \Uni_\Open^0
  }
  \mrow{
    \Tm : \Ty \to \Uni_\Open^0
  }
  \mrow{
    \gl{p}:\DecProp\ \ \text{(for $p\in\PP$)}
  }
  \mrow{
    \ExtConst_p : \prn{A : \Ty}\to\prn{a : \brc{\gl{p}}\, \Tm\,{A}} \to \Ty
  }
  \mrow{
    \Con{in}_p :
    \prn{A : \Ty}\,\prn{a : \brc{\gl{p}}\, \Tm\,{A}}
    \to
    \Ext{\Tm\,A}{\gl{p}}{a}
    \cong
    \Tm\,\prn{\ExtConst_p\,A\,a}
  }
}

\noindent
Following Remark~\ref{rem:normalization:isos}, we package the pair $\prn{\Con{in}_p,\Con{out}_p}$ as a
single isomorphism $\Con{in}_p$.

The presheaf of terms in the model $\AtmMod$ internalizes as a meta-space type
of \emph{variables} which by virtue of the structure map $\Mor{\AtmMod}{\IMod}$ can be indexed over
the object-space collection of terms. We realize this synthetically as follows:

\iblock{
  \mrow{\Vars : \prn{A:\Ty} \to \Ext{\Uni}{\Syn}{\Tm\,A}}
}

We refer to extensional type theory extended with these constants and modalities as the
language of \DefEmph{synthetic Tait computability} (STC).

\begin{remark}
  To account for strict universes---those for which $\ElConst$ commutes strictly with chosen
  codes---some prior STC developments employed strict gluing along the image of
  $\ElConst$~\citep{sterling:2021,sterling:phd}. By limiting our usage of strict gluing to $\Syn$, we
  are able to execute our constructions in a constructive metatheory. To model strict universes, we
  instead use the cumulativity of the hierarchy of universes $\Uni[i]$ and the fact that all
  levels are coherently closed under modalities and strict gluing.
\end{remark}

\subsection{Normal and neutral forms}
\label{sec:normalization:normals}

Internally to STC, we now specify the normal and neutral forms of terms, and
the normal forms of types. Following \citet{sterling:2021}
we index the type of neutral forms by a \DefEmph{frontier of instability}, a
proposition at which the neutral form is no longer meaningful.  Our
construction proceeds in two steps. First, we define a series of indexed
quotient-inductive definitions~\citep{kaposi:qiits:2019} specifying the
meta-space components of normal and neutral forms:

\iblock{
  \mrow{\Nf_\bullet : \prn{A : \Ty} \to \Tm\,A \to \Uni[\Closed]^0}
  \mrow{\Ne_\bullet : \prn{A : \Ty} \to \DecProp\to \Tm\,A \to \Uni[\Closed]^0}
  \mrow{\NfTy_\bullet : \Ty \to \Uni[\Closed]^0}
}

Next we use the strict gluing connective to define the types of normals, neutrals, and normal types
such that they lie strictly over $\Tm$ and $\Ty$:

\iblock{
  \mrow{\Nf\,A = \Glue{a}{\Tm\,A}{\NfAt{A}{a}}}
  \mrow{\Ne_\phi\,A = \Glue{a}{\Tm\,A}{\NeAt{A}{\phi}{a}}}
  \mrow{\NfTy = \Glue{A}{\Ty}{\NfTyAt{A}}}
}

\noindent
We illustrate a representative fragment of the inductive definitions in Figure~\ref{fig:normalization:rules}.

The induction principles for $\Nf_\bullet,\Ne_\bullet$ and $\NfTy_\bullet$ play
no role in the main development, which works with \emph{any} algebra for these
constants. These induction principles, however, are needed in order to prove
Theorem~\ref{thm:normalization:nf-decidable} and deduce the decidability of
definitional equality and the injectivity of type constructors. These same
considerations motivate our choice to index $\Ne_\bullet$ over $\DecProp$
rather than $\Prop$.

\begin{figure}[t]
  \begin{center}
    \iblock{
      \mrow{\NeVar : \prn{x : \Vars\,A} \to \NeAt{A}{\bot}{x}}
      \mrow{\NeUnstable : \brc{a : \Tm\,A} \to \NeAt{A}{\top}{a}}
      \mrow{\_ : \prn{a : \Tm\,A}\,\prn{e : \NeAt{A}{\top}{a}} \to e = \NeUnstable\,a}
      \row
      \mrow{\NfExt_p : \NfTyAt{A} \to \prn{\brc{p}\,\NfAt{A}{a}} \to \NfTyAt{\prn{\ExtConst_p\,A\,a}}}
      \mrow{\NfIn_p : \NfTyAt{A} \to \NfAt{A}{u} \to \NfAt{\prn{\ExtConst_p\,A\,a}}{u} }
      \mrow{
        \NeOut_p : \NfTyAt{A} \to \NeAt{\prn{\ExtConst_p\,A\,a}}{\phi}{u}
        \to \NeAt{A}{\prn{\phi\lor\gl{p}}}{\prn{\ExtIso_p\Inv\,u}}
      }
      \row
      \mrow{\NfNeU : \NeAt{\UniConst}{\phi}{A} \to \prn{\brc{\phi}\, \NfTyAt{\prn{\ElConst\,A}}} \to \NfTyAt{\prn{\ElConst\,A}}}
      \mrow{
        \NfNeNe : \NeAt{\UniConst}{\phi}{A} \to \prn{\brc{\phi}\, \NfTyAt\,\prn{\ElConst\,A}} \to \NeAt{\prn{\ElConst\,A}}{\psi}{a}
        \to \prn{\brc{\phi\lor\psi}\,\NfAt{\prn{\ElConst\,A}}{a}} \to \NfAt{\prn{\ElConst\,A}}{a}
      }
      \mrow{\NfNeEl : \NeAt{\UniConst}{\phi}{A} \to \prn{\brc{\phi}\, \NfAt{\UniConst}{A}} \to \NfAt{\UniConst}{A}}
      \row
      \mrow{\_ : \prn{e : \NeAt{\UniConst}{\top}{A}}\, \prn{u : \NfTyAt{A}} \to \NfNeU\,e\,u = u }
      \mrow{\_ : \prn{e : \NeAt{\UniConst}{\top}{A}}\, \prn{u : \NfAt{\UniConst}{A}} \to \NfNeEl\,e\,u = u}
      \mhang{
        \_ :
        \brc{\phi\lor\psi}\,
        \prn{e_A : \NeAt{\UniConst}{\phi}{A}}\,
        \prn{u_A : \brc{\phi}\,\NfTyAt{\prn{\ElConst\,A}}}
        \prn{e_a : \NeAt{\prn{\ElConst\,A}}{\psi}{a}}\,
        \prn{u_a : \NfAt{A}{a}}
      }{
        \mrow{
          \to \NfNeNe\,e_A\,u_A\,e_a\,u_a = u_a
        }
      }
    }
  \end{center}

  \caption{Selected rules from the definition of $\Nf$, $\Ne$, and $\NfTy$.}
  \label{fig:normalization:rules}
\end{figure}

\subsection{A glued normalization algebra}
\label{sec:normalization:algebra}

We can now construct a new \TTProp{}-algebra internally to $\Sh{\GlTop}$, satisfying the
constraint that each of its constituents restricts under $\Syn$ to the corresponding constant from
the \TTProp{}-algebra inherited from $\Sh{\ObjTop}$. We shall refer to this as the \DefEmph{normalization algebra}. For instance, we must define types representing
object types and terms:
\[
  \Ty* : \Ext{\Uni[2]}{\Syn}{\Ty}
  \qquad
  \Tm* : \Ext{\Ty* \to \Uni[1]}{\Syn}{\Tm}
\]

The meta-space component of the computability structure of types is given as a
dependent record below:

\iblock{
  \begin{make-rcd}{\Ty_\bullet\,\prn{A:\Ty}}{\Uni[2]}
    \mrow{\Code : \NfTyAt{A}}
    \mrow{\Pred : \Tm\,A \to \Uni[\Closed]^1}
    \mrow{
      \Reflect : \prn{a : \Tm\,A}\,\prn{\phi : \DecProp}\,\prn{e : \NeAt{A}{\phi}{a}}
      \to \prn{a_\phi : \brc{\phi}\,\Pred\,a} \to \Ext{\Pred\,a}{\phi}{a_\phi}
    }
    \mrow{\Reify : \prn{a : \Tm\,A} \to \Pred\,a \to \NfAt{A}{a}}
  \end{make-rcd}
}

The $\Pred$ field classifies the meta-space component of a given element; the
$\Reflect$ and $\Reify$ fields generalize the familiar operations of
normalization by evaluation, subject to Sterling and Angiuli's stabilization
yoga~\citep{sterling:2021}. We finally define both $\Ty*$ and $\Tm*$ using
strict gluing to achieve the correct boundary:

\iblock{
  \mrow{\Ty* = \Glue{A}{\Ty}{\Ty_\bullet\,A}}
  \mrow{\Tm*\,A = \Glue{a}{\Tm\,A}{\prn{\Unglue\,A}.\Pred\,a}}
}

\NewDocumentCommand\UnglDot{}{%
  \hspace{.1ex}%
  {\rotatebox{45}{\hbox{\rule{.3ex}{.3ex}}}}
  \hspace{.1ex}%
}

\begin{notation}
  Henceforth we will write $A\UnglDot\Con{fld}$ rather than
  $\prn{\Unglue{A}}.\Con{fld}$ to access a field of the closed component of
  $A$.
\end{notation}

We must also define $\gl{p}^* : \DecProp$ for each $p \in \mathbb{P}$ subject
to the condition that $\Syn$ implies $\gl{p}^* = \gl{p}$. As there is no
normalization data associated with these propositions, we define $\gl{p}^* =
\gl{p}$ which clearly satisfies the boundary condition.
It remains to show that $\prn{\Ty*,\Tm*}$ are closed under all the connectives of \TTProp. We show
two representative cases: extension types and the universe.

\subsubsection{\textbf{Extension types}}

Fixing $A : \Ty*$, $p : \mathbb{P}$, $a : \brc{\gl{p}}\,\Tm*\,A$, we must construct the
following pair of constants:

\iblock{
  \mrow{\ExtConst*_p\,A\,a : \Ext{\Ty*}{\Syn}{\ExtConst_p\,A\,a}}
  \mrow{
    \ExtIso*_p\,A\,a :
    \Ext{
      \Ext{\Tm*\,A}{\gl{p}}{a} \cong \Tm*\,\prn{\ExtConst*_p\,A\,a}
    }{\Syn}{
      \ExtIso_p\,A\,a
    }
  }
}

Recalling the definition of $\Ty*$ as a strict gluing type, we observe that the boundary condition
on $\ExtConst*_p$ already fully constrains the first component:

\iblock{
  \mrow{
    \ExtConst*_p\,A\,a = \MkGlue{\ExtConst_p\,A\,a}{
      \TypedHole{
        \Ty_\bullet\,\prn{\ExtConst_p\,A\,a}
      }
    }
  }
}

In the above, we have used Notation~\ref{not:normalization:mkglue} for constructing elements of a
strict gluing type.

We define the second component as follows, using copattern matching notation:

\iblock{
  \mrow{
    \prn{\ExtConst*_p\,A\,a}\UnglDot\Code = \NfExt_p\,A\UnglDot\Code\,\prn{A\UnglDot\Reify\,a}
  }
  \mrow{
    \prn{\ExtConst*_p\,A\,a}\UnglDot\Pred\,x = \Closed \Ext{A\UnglDot\Pred\,\prn{\ExtIso_p\Inv x}}{\gl{p}}{a}
  }
  \mrow{
    \prn{\ExtConst*_p\,A\,a}\UnglDot\Reify\,\prn{\eta_\bullet x} = \NfIn_p\,A\UnglDot\Code\,\prn{A\UnglDot\Reify\,x}
  }
  \mhang{
    \prn{\ExtConst*_p\,A\,a}\UnglDot\Reflect\,x\,\phi\,e\,\prn{\eta_\bullet x_{\phi}} =
  }{
    \mrow{
      \eta_\bullet
      \left(
      \begin{aligned}
          &A\UnglDot\Reflect\\
          &\ \ \prn{\ExtIso_p\Inv x} \,\prn{\phi \lor \gl{p}}
          \\
          &\ \ \prn{\NeOut_p\,A\UnglDot\Code\,e}\,\Split{\phi}{x_\phi}{\gl{p}}{a}
        \end{aligned}
      \right)
    }
  }
}

In the clauses of $\Reify$ and $\Reflect$, we were allowed to assume that the argument was of the
form $\eta_\bullet x$ where $\eta_\bullet$ is the unit of the modality $\Closed$:
$\eta_\bullet : A \to \Closed A$. This is because we are mapping into meta-space types and so this
``pattern-matching'' amounts to the bind operation of the monad $\Closed$.

\begin{remark}
  Stabilized neutrals are crucial to the definition of
  $\prn{\ExtConst*_p\,A\,a}\UnglDot\Reflect$ above: without them, we could not
  ensure that reflecting $\NeOut_p\,A\UnglDot\Code\,e$ lies within the
  specified subtype of $A\UnglDot\Pred$.
\end{remark}

The definition of $\ExtIso*_p$ is now straightforward:

\iblock{
  \mrow{\ExtIso*_p\,A\,a\,x = \MkGlue{\ExtIso_p\,A\,a\,x}{\Unglue\,x}}
}

We leave the routine verification of the various boundary conditions to the reader; nearly all of
them follow immediately from the properties of strict gluing.

\subsubsection{\textbf{The universe}}
We now turn to the construction of the universe in the normalization algebra; it is here that the
complexity of unstable neutrals becomes evident. Once again the boundary conditions on $\UniConst*$
force part of its definition:

\iblock{
  \mrow{\UniConst* = \MkGlue{\UniConst}{\TypedHole{\Ty_\bullet\,\UniConst}}}
}

The second component of $\UniConst*$ is complex and we present its definition in
Figure~\ref{fig:normalization:uni}. The inclusion of $\ElCode$ in $\UniConst_\bullet$ is necessary in
order to define $\ElConst*$:

\iblock{
  \mrow{\Syn \hookrightarrow \ElConst*\,A = \ElConst\,A}
  \mrow{\prn{\ElConst*\,\prn{\eta_\bullet A}}\UnglDot\Code = A\UnglDot\ElCode}
  \mrow{\prn{\ElConst*\,\prn{\eta_\bullet A}}\UnglDot\Pred = A\UnglDot\Pred}
  \mrow{\prn{\ElConst*\,\prn{\eta_\bullet A}}\UnglDot\Reflect = A\UnglDot\Reflect}
  \mrow{\prn{\ElConst*\,\prn{\eta_\bullet A}}\UnglDot\Reify = A\UnglDot\Reify}
}

Finally, we must show that $\UniConst*$ is closed under all small type formers and that $\ElConst*$
preserves them. This flows from the cumulativity of universes in $\Sh{\GlTop}$; to close
$\UniConst*$ under \eg{} products, we essentially `redo' the construction of products in $\Ty*$
by altering its predicate to be valued in $\Uni[0]$ rather than $\Uni[1]$.

\begin{figure}[t]
  \begin{center}
    \setlength\iblockleftmargin{0em}
    \raggedcolumns
    \iblock{
      \begin{make-rcd}{\UniConst_\bullet\,A}{\Uni[1]}
        \mrow{\Code:\NfAt{\UniConst}{A}}
        \mrow{\ElCode : \NfTyAt{\prn{\ElConst\,A}}}
        \mrow{\Pred : \Tm\,\prn{\ElConst\,A} \to \Uni[\Closed]^0}
        \mrow{
          \Reflect : \prn{a : \Tm\,A}\,\prn{\phi : \DecProp}
          \to \prn{e : \NeAt{A}{\phi}{a}}
          \to \prn{a_\phi : \brc{\phi}\,\Pred\,a}
          \to \Ext{\Pred\,a}{\phi}{a_\phi}
        }
        \mrow{\Reify : \prn{a : \Tm\,A} \to \Pred\,a \to \NfAt{A}{a}}
      \end{make-rcd}
    }

    \iblock{
      \mrow{\UniConst* : \Ext{\Ty*}{\Syn}{\UniConst}}
      \mrow{\UniConst*\UnglDot\Code = \NfUni}
      \mrow{\UniConst*\UnglDot\Pred\,A = \Closed\,\prn{\UniConst_\bullet\,A}}
      \mrow{\UniConst*\UnglDot\Reify\,\_\,\prn{\eta_\bullet A} = A.\Code}
    }

    \iblock{
      \mrow{
        \Syn \hookrightarrow \UniConst*\UnglDot\Reflect\,A\,\phi\,e_A\,A_\phi = A
      }
      \mhang{\Unglue\,\prn{\UniConst*\UnglDot\Reflect\,A\,\phi\,e_A\,A_\phi} =}{
        \mrow{\Kwd{let}\,A_\phi : \brc{\phi}\,\UniConst_\bullet\,A = X\gets A_\phi; X;}
        \mhang{\eta_\bullet\,\Kwd{record}}{
          \mrow{\Code =  \NfNeU\,e_A\,A_\phi\UnglDot\Code}
          \mrow{\ElCode = \NfNeEl\,e_A\,A_\phi\UnglDot\ElCode}

          \mrow{
            \Pred\,a =
            \Closed
            \prn{
              \prn{u : \NfAt{A}{x}} \times
              \brc{\phi}
              \Compr{
                a : A_\phi\UnglDot\Pred\,a
              }{
                A_\phi\UnglDot\Reify\,a = u
              }
            }
          }

          \mhang{\Reflect\,a\,\psi\,e_a\,a_\psi =}{
            \mrow{\Kwd{let}\,a_\psi : \brc{\psi}\,\prn{u : \NfAt{A}{x}}\times \ldots = x\leftarrow a_\psi; x;}
            \mrow{\Kwd{let}\, a_\phi = A_\phi\UnglDot\Reflect\,\psi\,e_a\,a_\psi;}
            \mrow{
              \eta_\bullet\prn{
                \NfNeEl\,e_A\,A_\phi\UnglDot\ElCode\, e_a\,\Split{\phi}{A_\phi\UnglDot\Reify\,a_\phi}{\psi}{\Proj{1}\,a_\psi}, a_\phi
              }
            }
          }
          \mrow{\Reify\,\_\,\prn{\eta_\bullet\,\prn{u,\_}} = u}
        }
      }
    }
  \end{center}

  \caption{The normalization structure on the universe.}
  \label{fig:normalization:uni}
\end{figure}

\subsubsection{The evaluation functor}\label{sec:normalization:evaluation-functor}

In this section, we equip $\Sh{\GlTop}$ with the maximal CwR structure, in which \emph{all} maps are representable.\footnote{It would be possible to choose a more restrictive class of representable maps for $\Sh{\GlTop}$, but there is no reason to do so.} We have just now defined an interpretation of \TTProp's signature (Section~\ref{sec:normalization:sig}) in $\Sh{\GlTop}$, and so by the universal property (Proposition~\ref{prop:normalization:tt-universal-prop}) of $\TT$ as the classifying CwR for this signature, we obtain a unique CwR functor $\Mor[I_{\Sh{\GlTop}}]{\TT}{\Sh{\GlTop}}$ sending every construct of \TTProp{} to its interpretation.

\subsection{The normalization algorithm}
\label{sec:normalization:algorithm}

Having constructed the normalization algebra in $\Sh{\GlTop}$ (Section~\ref{sec:normalization:algebra}), we can now define the actual normalization
function using an argument based on those presented by \citet[\S{}II.2]{fiore:2002,fiore:2022:nbe} and
\citet[\S{}3.3]{sterling:2025:grothendieck}, making use of the \emph{inserter model} of atomic substitutions introduced by \citet{bocquet-kaposi-sattler:2021}. As our results are constructive, our normalization function corresponds to an actual normalization by evaluation algorithm — whose executable computational content \citet{fiore:2022:nbe} has demonstrated explicitly in the simply typed case.

\subsubsection{Stripping of atomic contexts}\label{sec:normalization:stripping}

We first must establish an intermediate result: that the functor
$\Mor[\AtmMod\gl{-}]{\PP}{\AtmMod_\diamond}$ sending each $p\in\PP$ to the associated unary atomic
context is fully faithful and has a left adjoint, \ie that $\PP$ is reflective in
$\AtmMod_\diamond$. The left adjoint allows an atomic context to be ``stripped'' of anything that
induces variables, leaving only propositional assumptions. This result is ultimately used in
Lemma~\ref{lem:base-change-of-prf} to exhibit an isomorphism $\AtmMod\gl{p}\cong
\AtmFig^*\IMod\gl{p}$.

While it is straightforward to imagine how such a reflection can be defined by ``induction on atomic
contexts and repeated weakening'', we have not given an inductive specification of
$\AtmMod_{\diamond}$ and instead opted to specify it through its universal property. Accordingly, we
define this stripping map using an model to which we may apply the universal property of
$\AtmMod$. Fundamentally, however, the resulting constructions are the same, but our insistence on
using only these universal properties enables us to avoid fixing a particular and explicit
construction of atomic contexts.

\begin{construction}[The stripping model]
  We consider a model $\PMod$ of $\TT$ in which we set $\PMod_\diamond = \PP$,  $\PMod\prn{\Ty} = \PMod\prn{\Tm} = \ObjTerm{\Psh{\PP}}$, and $\PMod\gl{p} = \Yo{\PP}{p}$. All the remaining constructs of the model are trivial by virtue of these definitions. From the universal property of $\IMod$ as the bi-initial $\TT$-model, we obtain a unique homomorphism of models $\Mor[I_{\PMod}]{\IMod}{\PMod}$ whose contextual component is a product preserving functor $\Mor[I_{\PMod}^\diamond]{\IMod_\diamond}{\PP}$.
\end{construction}

\begin{lemma}
  The component $\Mor[\AtmMod\gl{-}]{\PP}{\AtmMod_\diamond}$ of the bi-initial atomic substitution model over $\IMod$ is full and faithful.
\end{lemma}

\begin{proof}
  Any functor out of a poset is necessarily faithful. To see that $\Mor[\AtmMod\gl{-}]{\PP}{\AtmMod_\diamond}$ is full, we fix a morphism $\Mor{\AtmMod\gl{p}}{\AtmMod\gl{q}}$ in $\AtmMod_\diamond$; as this morphism is necessarily unique, it suffices to show that $p\leq q$ in $\PP$. We consider the image of $\Mor{\AtmMod\gl{p}}{\AtmMod\gl{q}}$  under the contextual component of the composite homomorphism $\Mor[I_{\PMod}\circ\alpha]{\AtmMod}{\PMod}$ of $\TT_0$-models, which gives precisely the desired inequality $p\leq q$, recalling that each $\gl{r}$ is is representable in $\TT_0$ and thus preserved by homomorphisms of models.
\end{proof}

\begin{lemma}\label{lem:P-reflective-in-atcx}
  The functor between categories of contexts induced by
  $\Mor[I_{\PMod}\circ\alpha]{\AtmMod}{\PMod}$ is left adjoint to the embedding
  $\EmbMor[\AtmMod\gl{-}]{\PP}{\AtmMod_\diamond}$.
\end{lemma}

\begin{proof}
  The counit in $\PP$ is given by the identity inequality, as each $\gl{p}$ is representable in $\TT_0$ and thus preserved by homomorphisms. For the unit, we must construct a (necessarily unique) arrow $\Mor{\Gamma}{\AtmMod\gl{{I_{\PMod}^\diamond\prn{\alpha_\diamond \Gamma}}}}$ in $\AtmMod_\diamond$ for each atomic context $\Gamma$.

  For this, we consider a new atomic substitution model $\EMod$ over $\AtmMod$ whose category of
  contexts $\EMod_\diamond$ is the following inserter object~\citep[Section 6.5]{lack:2009} in $\CAT$:
  \[
    \begin{tikzpicture}[diagram]
      \node(E) {$\EMod_\diamond$};
      \node[right = of E] (A) {$\AtmMod_\diamond$};
      \node[right = 4cm of A] (A') {$\AtmMod_\diamond$};
      \draw[->,transform canvas={yshift=.5em}] (A) to node[above] {$1_{\AtmMod_\diamond}$} (A');
      \draw[->,transform canvas={yshift=-.5em}] (A) to node[below] {$\AtmMod\gl{-}\circ I_{\PMod}^\diamond\circ\alpha_\diamond$} (A');
      \draw[->,exists] (E) to node[above] {$\phi_\diamond$} (A);
    \end{tikzpicture}
  \]

  Equivalently, $\EMod_\diamond$ is the full subcategory of $\AtmMod_\diamond$ spanned by atomic contexts $\Gamma$ for which there exists an arrow $\Mor{\Gamma}{\AtmMod\gl{{I_{\PMod}^\diamond\prn{\alpha_\diamond \Gamma}}}}$; as the codomain is subterminal, such arrows are necessarily unique. We define all the constructs of $\TT_0$ in $\EMod$ as in $\AtmMod$, and it remains only to check that $\EMod$ has a terminal object and is closed under context comprehension and phase comprehension.

  \begin{enumerate}
    \item For the terminal object, we see that $\AtmMod\gl{{I_{\PMod}^\diamond\prn{\alpha_\diamond \ObjTerm{}}}}$ is already terminal.
    \item For the context comprehension, we fix $\Gamma\in\EMod_\diamond$ and $A\in\AtmMod\prn{\Ty}\prn{\Gamma}$, and we must check that there exists a map $\Mor{\Gamma.A}{\AtmMod\gl{{I_{\PMod}^\diamond\prn{\alpha_\diamond\prn{\Gamma.A}}}}}$. As $\alpha\circ I_\PMod$ is a homomorphism of models, it preserves context comprehensions; unraveling definitions, we ultimately have $I_{\PMod}^\diamond\prn{\alpha_\diamond\prn{\Gamma.A}}=I_{\PMod}^\diamond\prn{\alpha_\diamond\prn{\Gamma}}$ and so we are done.
    \item For phase comprehension, we fix $\Gamma\in\EMod_\diamond$ and $p\in\PP$ to check that there exists an arrow $\Mor{\Gamma.\AtmMod\gl{p}}{\AtmMod\gl{{I_{\PMod}^\diamond\prn{\alpha_\diamond\prn{\Gamma.\AtmMod{\prn{\gl{p}}}}}}}}$. But we have $I_{\PMod}^\diamond\prn{\alpha_\diamond\prn{\Gamma.\AtmMod{\prn{\gl{p}}}}} = I_{\PMod}^\diamond\prn{\alpha_\diamond\prn{\Gamma}}\land p$, so we may use the projection $\Mor{\Gamma.\AtmMod\gl{p}}{\AtmMod\gl{p}}$.
  \end{enumerate}

  We evidently have a homomorphism of $\TT_0$-models $\Mor[\eta]{\EMod}{\AtmMod}$ that exhibits
  $\EMod$ as a atomic substitution model over $\AtmMod$. Postcomposing with the structure map
  $\Mor[\alpha]{\AtmMod}{\IMod}$, we can view $\EMod$ as a atomic substitution model over $\IMod$.
  Thus, by the universal property of $\AtmMod$ we have a universal \emph{section}
  $\Mor[J_{\EMod}]{\AtmMod}{\EMod}$ to $\Mor[\eta]{\EMod}{\AtmMod}$. This shows that every atomic
  context $\Gamma\in\AtmMod_\diamond$ can be equipped with an arrow
  $\Mor{\Gamma}{\AtmMod\gl{{I_{\PMod}^\diamond\prn{\alpha_\diamond \Gamma}}}}$. Assembling all these
  arrows together, we have the unit of the adjunction $I_{\PMod}^\diamond\circ\alpha_\diamond\dashv
  \AtmMod\gl{-}$.
\end{proof}

The force of Lemma~\ref{lem:P-reflective-in-atcx} is to show that $\PP$ is a reflective subcategory of $\AtmMod_\diamond$.

\subsubsection{Computability spaces of atomic and computable substitutions}

We will consider two computability spaces induced by an atomic context $\Gamma$:  the computability space $\bbrk{\Gamma}$ of ``computable substitutions into $\Gamma$'' and the computability space $\pprn{\Gamma}$ of ``atomic substitutions into $\Gamma$''.

\begin{construction}[The computability space of computable substitutions]
  The computability space $\bbrk{\Gamma}$ of computable substitutions into an atomic context $\Gamma$ is defined in terms of the \emph{interpretation} of $\TT$ into $\Sh{\GlTop}$ as follows, sending each atomic context to the computability space determined by the algebra structure:
  \[
    \begin{tikzpicture}[diagram]
      \node (0) {$\AtmMod_\diamond$};
      \node[right = of 0] (1) {$\IMod_\diamond$};
      \node[right = of 1] (2) {$\TT$};
      \node[right = of 2] (3) {$\Sh{\GlTop}$};
      \draw[->] (0) to node[above] {$\alpha_\diamond$} (1);
      \draw[embedding] (1) to node[above] {$\subseteq$} (2);
      \draw[->] (2) to node[above] {$I_{\Sh{\GlTop}}$} (3);
      \draw[->, exists,bend right=30] (0) to node[below] {$\bbrk{-}$} (3);
    \end{tikzpicture}
  \]
\end{construction}

\begin{construction}[The computability space of atomic substitutions]
  We  define an embedding $\EmbMor[\pprn{-}]{\AtmMod_\diamond}{\Sh{\GlTop}}$ sending $\Gamma\in\AtmMod_\diamond$ to the computability space $\pprn{\Gamma}$ with $\OpenIncl^*\pprn{\Gamma} = \Yo{\IMod_\diamond}{\alpha_\diamond\Gamma}$ and $\ClIncl^*\pprn{\Gamma} = \Yo{\AtmMod_\diamond}\Gamma$, such that $\Mor[\pi_{\pprn{\Gamma}}]{\Yo{\AtmMod_\diamond}\Gamma}{\AtmFig^*\Yo{\IMod_\diamond}\alpha_\diamond\Gamma}$ is defined on generalised elements by the functorial action of $\Mor[\alpha_\diamond]{\AtmMod_\diamond}{\IMod_\diamond}$ as follows:
  \[
    \pi_{\pprn{\Gamma}}^\Delta\prn{\Mor[\gamma]{\Delta}{\Gamma}} =
    \Mor[\alpha_\diamond{\gamma}]{\alpha_\diamond \Delta}{\alpha_\diamond\Gamma}
  \]

  Miraculously, the \emph{coreflective} embedding $\EmbMor[\ClIncl_!]{\Sh{\MetaTop}}{\Sh{\GlTop}}$ sends $\Yo{\AtmMod_\diamond}\Gamma$ to \textbf{precisely} the computability space $\pprn{\Gamma}$, up to isomorphism:
  \begin{align*}
    \ClIncl_!\Yo{\AtmMod_\diamond}\Gamma
    &=
    \prn{\AtmFig_!\Yo{\AtmMod_\diamond}\Gamma, \Mor{\Yo{\AtmMod_\diamond}\Gamma}{\AtmFig^*\AtmFig_!\Yo{\AtmMod_\diamond}\Gamma}}
    \\
    &\cong
    \prn{\Yo{\IMod_\diamond}\alpha_\diamond\Gamma, \Mor{\Yo{\AtmMod_\diamond}\Gamma}{\AtmFig^*\Yo{\IMod_\diamond}\alpha_\diamond\Gamma}}
    \\
    &=
    \pprn{\Gamma}
  \end{align*}

\end{construction}

Note that the functors $\bbrk{-},\pprn{-}$ lift into the slice $\CAT\downarrow\Sh{\ObjTop}$ in the following sense:
\[
  \begin{tikzpicture}[diagram]
    \node (A) {$\AtmMod_\diamond$};
    \node[left = of A] (G/l) {$\Sh{\GlTop}$};
    \node[right = of A] (G/r) {$\Sh{\GlTop}$};
    \node[below = of A] (ShI) {$\Sh{\ObjTop}$};
    \draw[->] (A) to node[upright desc] {$\Yo{\IMod_\diamond}\circ \alpha_\diamond$} (ShI);
    \draw[->] (A) to node[above] {$\bbrk{-}$} (G/r);
    \draw[->] (A) to node[above] {$\pprn{-}$} (G/l);
    \draw[->] (G/l) to node[sloped,below] {$\OpenIncl^*$} (ShI);
    \draw[->] (G/r) to node[sloped,below] {$\OpenIncl^*$} (ShI);
  \end{tikzpicture}
\]

\begin{notation}
  Let $\Gamma$ be an atomic context, and let $\Mor[A]{\pprn{\Gamma}}{\Ty}$ be an object-space type, which we may regard as a morphism $\Mor{\alpha\Gamma}{\Ty}$ in $\TT$. We shall write $\bbrk{A}\colon \bbrk{\Gamma}\to \Ty*$ for the image of $A$ under the interpretation functor $I_{\Sh{\GlTop}}$.
\end{notation}

\begin{lemma}\label{lem:atomic-and-canonical-cx-ext}
  Let $\Gamma$ be an atomic context, and let $\Mor[A]{\pprn{\Gamma}}{\Ty}$ be an object-space type
  (which we may regard as a morphism $\Mor{\alpha_\diamond\Gamma}{\Ty}$ in $\TT$). Then we have the following cartesian squares:
  \[
    \DiagramSquare{
      se = \Ty,
      ne = \Sum{A:\Ty}\Vars\, A,
      sw = \pprn{\Gamma},
      south = A,
      nw = \pprn{\Gamma.A},
      nw/style = pullback,
      east = \pi_1,
      west = \pprn{p_A},
    }
    \qquad
    \DiagramSquare{
      nw/style = pullback,
      se = \Ty*,
      ne = \Sum{A:\Ty*}\Tm*{A},
      sw = \bbrk{\Gamma},
      nw = \bbrk{\Gamma.A},
      east = \pi_1,
      south = \bbrk{A},
    }
  \]

  Stated in the internal language, we have canonical isomorphisms $\pprn{\Gamma.A}\cong \Sum{\gamma:\pprn{\Gamma}}\Vars\,\prn{A\gamma}$ and $\bbrk{\Gamma.A} \cong \Sum{\gamma:\bbrk{\Gamma}}\Tm*\,\prn{\bbrk{A}\gamma}$.
\end{lemma}

\begin{proof}
  The latter is the image of a pullback square in $\TT$ under $I_{\Sh{\GlTop}}$, which is finitely continuous. The former can be seen by means of an explicit computation.
\end{proof}

We now come to an important result relating the interpretation of $\gl{p}$ in $\IMod$ to the interpretation of the same in $\AtmMod$. Lemma~\ref{lem:base-change-of-prf} below is the \emph{raison d'\^etre} for the stripping model $\PMod$ in Section~\ref{sec:normalization:stripping}.

\begin{lemma}\label{lem:base-change-of-prf}
  We have a (necessarily unique) isomorphism $\AtmMod\gl{p}\cong \AtmFig^*\IMod\gl{p}$.
\end{lemma}

\begin{proof}
  As both presheaves are subterminal, it is enough to see that one is inhabited if and only if the other is.
  \begin{enumerate}
    \item We may transpose a map $\Mor{\Yo{\AtmMod_\diamond}\Gamma}{\AtmFig^*\IMod\gl{p}}$ to get $\Mor{\alpha_\diamond\Gamma}{\IMod\gl{p}}$; applying the functorial action of $\Mor[I_{\PMod}^\diamond]{\IMod_\diamond}{\PP}$, we have $I_{\PMod}^\diamond\alpha_\diamond\Gamma\leq I_{\PMod}^\diamond\IMod\gl{p}=p$ and by adjoint transpose with Lemma~\ref{lem:P-reflective-in-atcx}, we have $\Mor{\Gamma}{\AtmMod\gl{p}}$.
    \item Conversely, given an arrow $\Mor{\Gamma}{\AtmMod\gl{p}}$ we may apply the functorial action of $\Mor[\alpha_\diamond]{\AtmMod_\diamond}{\IMod_\diamond}$ to obtain an arrow $\Mor{\alpha_\diamond\Gamma}{\alpha_\diamond\AtmMod\gl{p}}$. As $\gl{p}$ is representable in $\TT_0$, it is preserved by morphisms of models like $\Mor[\alpha]{\AtmMod}{\IMod}$; thus $\alpha_\diamond\AtmMod\gl{p}\cong \IMod\gl{p}$ and so we have $\Mor{\alpha_\diamond\Gamma}{\IMod\gl{p}}$, which we may transpose to obtain $\Mor{\Yo{\AtmMod_\diamond}\Gamma}{\AtmFig^*\IMod\gl{p}}$. \qedhere
  \end{enumerate}
\end{proof}

\subsubsection{Hydration of atomic substitutions}

A critical point in concrete normalization by evaluation algorithms is to ``reflect'' a vector of variables as an environment of (computable) values against which the computability interpretation of an open term can be executed; in a concrete setting, this operation is defined by recursion on the atomic contexts. The same process, which we shall refer to here as the \DefEmph{hydration of atomic substitutions}, plays an equally important role in semantic proofs of normalization in the guise of a certain \DefEmph{hydration map} $\Mor[\nearrow]{\pprn{-}}{\bbrk{-}}$ in $\CAT\downarrow\Sh{\ObjTop}$ that we shall need to construct.

Just as in the definition of the stripping map, we are confronted by the fact that
 we have defined the atomic contexts only by means of a universal property (the bi-initial atomic substitution model over $\IMod$), so we do not immediately have anything concrete to do recursion on. The innovation of \citet{bocquet-kaposi-sattler:2021} was to find the correct \emph{categorical induction motive} that explains the usual recursive argument purely in terms of the universal property of the bi-initial atomic substitution model (likewise due to \opcit). In what follows, we adapt their ideas to our setting and show how to construct the desired hydration map.

We begin by defining a new model $\HMod$ of $\TT_0$ that we shall refer to as the \DefEmph{hydration model}. The category of contexts $\HMod_\diamond$ is defined to be the underlying category of the \emph{inserter object} for $\Mor[\pprn{-},\bbrk{-}]{\AtmMod_\diamond}{\Sh{\GlTop}}$ in $\CAT\downarrow\Sh{\ObjTop}$ as depicted below:
\[
  \begin{tikzpicture}[diagram]
    \node(H) {$\HMod_\diamond$};
    \node[right = of H] (A) {$\AtmMod_\diamond$};
    \node[right = of A] (G) {$\Sh{\GlTop}$};
    \node[below = of A] (I) {$\Sh{\ObjTop}$};
    \draw[embedding,transform canvas={yshift=.5em}] (A) to node[above] {$\pprn{-}$} (G);
    \draw[->,transform canvas={yshift=-.5em}] (A) to node[below] {$\bbrk{-}$} (G);
    \draw[->,exists] (H) to (I);
    \draw[->,exists] (H) to node[above] {$\psi_\diamond$} (A);
    \draw[->] (A) to node[upright desc] {$\Yo{\IMod_\diamond}\circ\alpha_\diamond$} (I);
    \draw[->] (G) to node[sloped,below] {$\OpenIncl^*$} (I);
  \end{tikzpicture}
\]

Explicitly, an object of $\HMod_\diamond$ is a pair $\prn{\Gamma,h_\Gamma}$ of an object $\Gamma\in \AtmMod_\diamond$ together with an arrow $\Mor[h_\Gamma]{\pprn{\Gamma}}{\bbrk{\Gamma}}$ whose image under $\Mor[\OpenIncl^*]{\Sh{\GlTop}}{\Sh{\ObjTop}}$ is the identity map on $\Yo{\IMod_\diamond}\alpha_\diamond\Gamma$. An arrow from $\prn{\Delta,h_\Delta}$ to $\prn{\Gamma,h_\Gamma}$ is given by an arrow $\Mor[\gamma]{\Delta}{\Gamma}$ in $\AtmMod_\diamond$ making the following square commute:
\[
  \DiagramSquare{
    nw = \pprn{\Delta},
    sw = \bbrk{\Delta},
    ne = \pprn{\Gamma},
    se = \bbrk{\Gamma},
    west = h_\Delta,
    east = h_\Gamma,
    north = \pprn{\gamma},
    south = \bbrk{\gamma},
  }
\]

Clearly, $\HMod_\diamond$ has a terminal object because $\bbrk{\ObjTerm{\AtmMod_\diamond}}$ is terminal. Anticipating that $\psi_\diamond$ should lift to a morphism of models exhibiting $\HMod$ as a atomic substitution model over $\AtmMod$, we define $\HMod\gl{p} = \psi_\diamond^*\AtmMod\gl{p}$ and $\HMod\prn{\Ty} = \psi_\diamond^*\AtmMod\prn{\Ty}$. In order to define $\HMod\prn{\Tm}$, it will be easiest to first define context comprehensions.

\begin{construction}[Context comprehensions in $\HMod_\diamond$]
  Given a context $\prn{\Gamma,h_\Gamma}$ in $\HMod_\diamond$ and a type $\Mor[A]{\Yo{\AtmMod_\diamond}\Gamma}{\AtmMod\prn{\Ty}}$, we can lift the context comprehension $\Gamma.A\in \AtmMod_\diamond$ into $\HMod_\diamond$ by finding a suitable map $\Mor[h_{\Gamma.A}]{\pprn{\Gamma.A}}{\bbrk{\Gamma.A}}$ whose image in $\Sh{\ObjTop}$ is the identity. Recalling Lemma~\ref{lem:atomic-and-canonical-cx-ext}, we may equivalently construct (from the internal point of view) the following map:

  \iblock{
    \mrow{
      h_{\Gamma.A} \colon
      \Ext{
        \prn{\Sum{\gamma:\pprn{\Gamma}}\Vars\,\prn{A\gamma}}
        \to
        \Sum{\gamma:\bbrk{\Gamma}}\Tm*\prn{\bbrk{A}\gamma}
      }{\Syn}{\lambda u\pdot u}
    }
    \mrow{
      h_{\Gamma.A}\prn{\gamma,x} =
      \prn{
        h_\Gamma\gamma,
        \bbrk{A}\prn{h_\Gamma\gamma}\UnglDot
        \Reflect\,x\,\bot\,\prn{\NeVar\,x}\,\prn{\lambda\_:\bot\pdot \star}
      }
    }
  }

  The projection $\Mor[p_A]{\Gamma.A}{\Gamma}$ tracks a morphism $\Mor{\prn{\Gamma.A,h_{\Gamma.A}}}{\prn{\Gamma,h_\Gamma}}$, by definition of $h_{\Gamma.A}$.
\end{construction}

\begin{construction}[Phase comprehensions in $\HMod_\diamond$]
  Given $\Gamma\in \HMod_\diamond$ and $p\in \PP$, we must exhibit a map $\Mor[h_{\Gamma.\AtmMod\gl{p}}]{\pprn{\Gamma.\AtmMod\gl{p}}}{\bbrk{\Gamma.\AtmMod\gl{p}}}$. Using Lemma~\ref{lem:atomic-and-canonical-cx-ext}, we construct this by combining the assumed map $\Mor[h_\Gamma]{\pprn{\Gamma}}{\bbrk{\Gamma}}$ with the (necessarily unique) map $\Mor{\pprn{\AtmMod\gl{p}}}{\bbrk{\AtmMod\gl{p}}}$ obtained from the identity map on $\AtmMod\gl{p}$ in the following way:
  \begin{enumerate}
    \item First we observe that $\bbrk{\AtmMod\gl{p}} \cong \OpenIncl_*\IMod\gl{p}$ as follows:
    \begin{align*}
      \bbrk{\AtmMod\gl{p}}
      &=
      I_{\Sh{\GlTop}}\alpha_\diamond\AtmMod\gl{p}
      &&\text{by definition}
      \\
      &\cong
      I_{\Sh{\GlTop}}\IMod\gl{p}
      &&\text{$\alpha$ is a homomorphism}
      \\
      &\cong
      \OpenIncl_*\IMod\gl{p}
      &&\text{by definition}
    \end{align*}
    \item Then we proceed by adjoint calisthenics:
      \begin{align*}
        \Hom{\Sh{\GlTop}}{\pprn{\AtmMod\gl{p}}}{\bbrk{\AtmMod\gl{p}}}
        &\cong
        \Hom{\Sh{\GlTop}}{\pprn{\AtmMod\gl{p}}}{\OpenIncl_*\IMod\gl{p}}
        &&\text{by the above}
        \\
        &\cong
        \Hom{\Sh{\ObjTop}}{\OpenIncl^*\pprn{\AtmMod\gl{p}}}{\IMod\gl{p}}
        &&\text{by adjoint transpose}
        \\
        &\cong
        \Hom{\Sh{\ObjTop}}{\AtmFig_!\AtmMod{\gl{p}}}{\IMod{\gl{p}}}
        &&\text{by definition of $\pprn{-}$}
        \\
        &\cong
        \Hom{\Sh{\MetaTop}}{\AtmMod{\gl{p}}}{\AtmFig^*\IMod{\gl{p}}}
        &&\text{by adjoint transpose}
        \\
        &\cong
        \Hom{\Sh{\MetaTop}}{\AtmMod{\gl{p}}}{\AtmMod\gl{p}}
        &&\text{by Lemma~\ref{lem:base-change-of-prf}}
        \qedhere
      \end{align*}
  \end{enumerate}
\end{construction}

\begin{construction}[The presheaf of terms]
  We define $\HMod\prn{\Tm}\in \Psh{\HMod_\diamond}\downarrow \HMod\prn{\Ty}$ to send $A\in \HMod\prn{\Ty}\prn{\Gamma,h_\Gamma}$ to the set of sections of the projection $\Mor[p_A]{\prn{\Gamma.A,h_{\Gamma.A}}}{\prn{\Gamma,h_\Gamma}}$ in $\HMod_\diamond$.
\end{construction}

\begin{construction}[The hydration model]
  With the definitions that we have given, the projection functor $\Mor[\psi_\diamond]{\HMod_\diamond}{\AtmMod_\diamond}$ easily extends to a morphism of models, exhibiting $\HMod$ as a atomic substitution model over $\AtmMod$.
\end{construction}

\begin{construction}[The hydration map]\label{con:hydration}
  As we may compose $\Mor[\psi]{\HMod}{\AtmMod}$ with the structure map $\Mor[\alpha]{\AtmMod}{\IMod}$, we can view $\HMod$ as a atomic substitution model over $\IMod$ as well, and thus by the universal property of $\AtmMod$ we have a universal section $\Mor{\AtmMod}{\HMod}$ to $\Mor[\psi]{\HMod}{\AtmMod}$. Unraveling the section $\Mor{\AtmMod}{\HMod}$ we obtain precisely a natural assignment of hydration map component $\Mor[h_\Gamma]{\pprn{\Gamma}}{\bbrk{\Gamma}}$ from which we may assemble a single hydration map $\Mor[\nearrow]{\pprn{-}}{\bbrk{-}}$ in $\CAT\downarrow\Sh{\ObjTop}$.
\end{construction}

\subsubsection{Normalization and decidability}\label{sec:normalizing-and decidability}

We can now show how to compute the normal form of a type $\IsTy[\Gamma]{A}$, which we regard as an arrow $\Mor[A]{\alpha_\diamond\Gamma}{\Ty}$ in $\TT$. Then we may apply the functorial action of the evaluation functor $\Mor[I_{\Sh{\GlTop}}]{\TT}{\Sh{\GlTop}}$ to obtain $\Mor[\bbrk{A}]{\bbrk{\Gamma}}{\Ty*}$ and then postcompose with the projection of normal forms to obtain $\Mor[\bbrk{A}\UnglDot\Code]{\bbrk{\Gamma}}{\NfTy}$. Unraveling the meaning of such a map in the gluing category $\Sh{\GlTop}$, we see that this amounts \emph{not} to a normal form for $A$ but instead to an assignment of normal forms of $A$ to computability witnesses for all the variables in the context $\Gamma$. It is precisely this gap that hydration fills:
\[
  \mathsf{norm}_{\Ty}\prn{\IsTy[\Gamma]{A}} =
  \pprn{\Gamma}\xrightarrow{\nearrow_\Gamma}\bbrk{\Gamma}\xrightarrow{\bbrk{A}}\Ty*\xrightarrow{-\UnglDot\Code}\NfTy
\]
The entire map above restricts within object-space, by construction, to the original type $A$, or (to be more precise) its image in $\Sh{\GlTop}$ under $\EmbMor[\OpenIncl_*]{\Sh{\ObjTop}}{\Sh{\GlTop}}$. The meta-space component of such a morphism $\Mor{\pprn{\Gamma}}{\NfTy}$ is precisely a normal form for $A$.

\begin{therm}
  Normalization is sound and complete:
  \begin{enumerate}
    \item \textbf{Soundness.} If $\mathsf{norm}_{\Ty}\prn{\IsTy[\Gamma]{A}} = \mathsf{norm}_{\Ty}\prn{\IsTy[\Gamma]{B}}$, then $\EqTy[\Gamma]{A}{B}$.
    \item \textbf{Completeness.} If $\EqTy[\Gamma]{A}{B}$, then $\mathsf{norm}_{\Ty}\prn{\IsTy[\Gamma]{A}} = \mathsf{norm}_{\Ty}\prn{\IsTy[\Gamma]{B}}$.
  \end{enumerate}
\end{therm}

\begin{proof}
  Completeness holds \emph{by definition}, as the normalization function is defined on the denizens of the syntactic CwR rather than on raw terms. Soundness follows from the fact that $\mathsf{norm}_{\Ty}\prn{\IsTy[\Gamma]{A}}$ restricts in object-space to $A$ itself.
\end{proof}

In the same way, we can construct a normalization function for terms and prove that it is sound and complete (though we do not do so here). Of course, deciding equality for terms is practically important only insofar as it arises in the context of deciding equality for the types that mention them.

\begin{definition}
  An object $X \in \Sh{\GlTop}$ has \DefEmph{levelwise decidable equality} when
  for each $\Gamma \in \AtmMod\Sub{\diamond}$, the set $\prn{\ClIncl^*X}\Gamma$
  has decidable equality where $\Mor|closed immersion|[\ClIncl]{\MetaTop}{\GlTop}$ is as
  in Definition~\ref{def:gluing-topos}.
\end{definition}

\begin{therm}
  \label{thm:normalization:nf-decidable}
  Viewed as objects of $\Sh{\GlTop}$, the following have levelwise decidable equality:
  \[
    \NfTy
    \qquad
    \prn{A : \Ty} \times \Nf\,A
    \qquad
    \prn{A : \Ty} \times \prn{\phi:\DecProp} \times \Ne_\phi{A}
  \]
\end{therm}

From this, we obtain our main results concerning \TTProp:

\begin{corollary}
  Definitional equality %
  in \TTProp{} is decidable.
\end{corollary}

\subsubsection{Stronger normalization results}

A few stronger results can be proved using routine extensions of the methods on display here.

\begin{enumerate}
  \item \emph{The external normalization function is surjective, which implies that normalization is idempotent.} The main practical impact of normalization being surjective is to prove that the normalization function is effectively computable, as in \citet{sterling:phd,sterling:2021}; this step is redundant in the setting of the present paper, which has been carried out constructively in order to ensure an implicit form of effective computability.

  \item \emph{Type constructors are injective in the sense that $\EqTy{A\to B}{A'\to B'}$ implies $\EqTy{A}{A'}$ and $\EqTy{B}{B'}$, \etc.} Injectivity of type constructors is the main ingredient to establishing determinacy of the standard \emph{bidirectional elaboration} algorithm. Injectivity is not strictly needed for a type checker written on fully annotated terms, but practical systems involve the elaboration of less-annotated terms to fully annotated terms; this process relies on injectivity.

  \item \emph{Normalization can be internalised into $\Sh{\GlTop}$ as an inverse $\Mor{\Ty}{\NfTy}$ to the projection $\NfTy\to \Ty$, as in \citet{sterling:phd,sterling:2025:grothendieck,sterling:2021}.} This implies, for example, that the normalization function is invariant under variable renaming (or, more generally, atomic substitutions).
\end{enumerate}

We do not detail these results here, but instead remark that detailed proofs for similar theories can be found in the cited literature.

\section{Related work}
\label{sec:related-work}

Proof assistants already have support for various means of controlling the unfolding of definitions;
we classify these as either \emph{library-} or \emph{language-level}.

\subsection{Library-level features}
Various library-level idioms for abstract definitions are used in practice such
as \textsc{SSReflect}'s \texttt{lock} idiom. While such approaches are flexible
and compatible with existing proof assistants, they are often cumbersome in
practice. For instance, \texttt{lock} relies on various tactics with subtle
behavior, which makes it difficult to use \texttt{lock}ing idioms in pure
Gallina code.

\subsection{Language-level features}
Many proof assistants include a feature like Agda's \texttt{abstract} blocks which marks a
definition as completely opaque to the remainder of the development. In
Remark~\ref{rem:unfolding-sections}, we explained how to recover Agda's \texttt{abstract} definitions
using controlled unfolding. Moreover, as controlled unfolding does not require a user to decide
up front whether a definition can be unfolded, it gives a more realistic and flexible discipline
for abstraction in a proof assistant.
In practice, however, \texttt{abstract} is often used for performance reasons instead of merely for
controlling abstraction; unfolding large or complex definitions can significantly slow down type
checking and unification. While we have not discussed performance considerations for controlled
unfolding, the same optimizations apply to our mechanism for definitions that are never
unfolded.
In total, controlled unfolding strictly generalizes \texttt{abstract} blocks.

Recently, Kov{\'a}cs~\citep{smalltt,kovacs:2024} has proposed a \emph{glued evaluation} technique to
both improve the pretty-printing of goals and more efficiently handle unfolding during conversion
testing. Roughly, the proof assistant's kernel may choose to unfold any definition, but avoids doing
so whenever possible for efficiency and strives to never show unfolded goals to the user. Both glued
evaluation and controlled unfolding relate to the unfolding of definitions, they are largely
orthogonal and complementary. In particular, glued evaluation does not require user intervention,
unlike controlled unfolding, but it does not actually preclude any unfolding from taking
place. Thus, glued evaluation does not impact the well-formedness of a program and can be used as a
``drop-in'' technique for improving performance and usability. However, for the same reasons, glued
evaluation cannot be used to enforce modularity and independence in the same way as controlled
unfolding. Ideally, a proof assistant would support both glued evaluation and controlled unfolding:
the more advanced evaluation algorithm would improve baseline performance and controlled unfolding
would facilitate users enforcing stronger abstraction boundaries within their programs and assisting
the kernel by manually designating certain definitions as opaque.

Program verifiers such as VeriFast and Chalice include similar unfolding mechanisms to cope
specifically with recursive definitions~\citep{jacobs:2015,summers:2013}. Like our mechanism, these
features allow users fine-grained control over how definitions are unfolded. However, these
verifiers work only within simply-typed theories and thus avoid the substantial complexity of
dependency. Moreover, these mechanisms manage a different problem than controlled unfolding; they
allow a user to unfold recursive definitions step-by-step while controlled unfolding is used to
control when each definition can be fully inlined.

\subsection{Translucent ascription in module systems}
Thus far we have focused on proof assistants, but similar considerations arise
for ML-style module
systems~\citep{milner:1997,harper:2000,dreyer:2003,sterling:modules:2021}.
The default opacity for definitions in module systems is the
same as in controlled unfolding and opposite to proof assistants: types are
abstract unless marked otherwise.
The treatment of translucent type declarations in module
systems~\citep{harper:2000} relies on \emph{singleton
kinds}~\citep{aspinall:1995,stone-harper:2006}, which are the special case of
extension types whose boundary proposition is $\top$.
Generalizing from compiletime kinds to mixed compiletime--runtime \emph{module signatures}, Sterling
and Harper have pointed out that transparent ascriptions are best handled by an extension type whose
boundary proposition represents the compiletime phase itself~\citep{sterling:modules:2021}. Thus the
translucency of compiletime module components can be seen as a \emph{particular} controlled
unfolding policy in the sense of this paper.

\subsection{Controlled unfolding in Agda}

Inspired by our implementation of controlled unfolding in \cooltt, Am{\'e}lia Liao and Jesper Cockx
have implemented a version of this mechanism called \texttt{opaque} within Agda
2.6.4~\citep{agda-unfolding:2022}. However, rather than using extension types, their Agda
implementation simulates the necessary behaviors by instrumenting conversion checking---a workaround
made possible by the very restricted ways in which our elaboration procedure relies on extension
types. This demonstrates that controlled unfolding can be adapted to proof assistants like Coq whose
core calculi do not presently support extension types.

At the time of writing, Agda's \texttt{opaque} declarations are new enough that only two major Agda
libraries, the 1Lab \citep{1lab} and the Cubical Agda library \citep{agda-cubical}, use them
extensively; the Agda standard library may adopt \texttt{opaque} declarations in a future major
revision \citep{agda-stdlib-opaque}. As of publication, 65 modules in the 1Lab use \texttt{opaque}
and 19 use \texttt{unfolding}, in addition to over 100 using \texttt{abstract} blocks; in the
Cubical Agda library, 27 modules use \texttt{opaque}, 6 use \texttt{unfolding}, and 35 use
\texttt{abstract}.

\section{Conclusions and future work}
\label{sec:conclusions}

We have proposed \emph{controlled unfolding}, a new mechanism for interpolating between transparent
and opaque definitions in proof assistants. We have demonstrated its practical applicability by
extending \cooltt{} with controlled unfolding; we have also proved its soundness through an
elaboration algorithm to a core calculus whose normalization we establish using a constructive
synthetic Tait computability argument.

In the future, we hope to see controlled unfolding integrated into more proof assistants and to
further explore its applications for large-scale organization of mechanized mathematics. As
mentioned above, some our mechanism has implemented in Agda, but features such as local unfolds are
still absent. Furthermore, in the context of our \cooltt{} implementation, we have also already
begun to experiment with potential extensions, including one that allows a \emph{subterm} to be
declared locally abstract and then unfolded later on as-needed --- a more flexible alternative to
Coq's $\Prg{abstract}~t$ tactical. As we mentioned in Remark~\ref{rem:unfolding-sections}, we also are
interested in facilities to limit the scope in which it is possible to unfold a definition.

\section*{Acknowledgments}

This work was supported in part by a Villum Investigator grant (no. 25804), Center for Basic
Research in Program Verification (CPV), from the VILLUM Foundation.
Jonathan Sterling is funded by the European Union under the Marie Sk\l{}odowska-Curie Actions
Postdoctoral Fellowship project \emph{TypeSynth: synthetic methods in program verification},
and by the United States Air Force Office of Scientific Research under grant number
FA9550-23-1-0728 (\emph{New Spaces for Denotational Semantics}; Dr.\ Tristan Nguyen, Program
Manager).
Carlo Angiuli is supported by the U.S.\ Air Force Office of Scientific Research under grant
numbers FA9550-21-1-0009 and FA9550-24-1-0350.
Any opinions, findings and conclusions or recommendations expressed in this material are
those of the authors and do not necessarily reflect the views of the European Union, the
European Commission, or the AFOSR. Neither the European Union nor the granting authority can
be held responsible for them.

\section*{Competing interests}
The authors declare no competing interests.

\printbibliography

\clearpage
\appendix
\section{The (2,1)-category of models}
\label{app:category-of-models}

\citet{uemura:phd} has observed that a model
$\prn{\MMod_\diamond,\MMod}$ in the sense of Definition~\ref{def:taichi-model} can be packaged
into a single functor $\Mor[\Tot{\MMod}]{\TT^\rhd}{\CAT}$, in which
$\TT^\rhd$ freely extends $\TT$ by a new terminal object $\diamond$ and $\CAT$
is the 2-category of categories. From this perspective, a sort $X\in\TT$ is
taken to the total category $\Tot{\MMod}\prn{X} =
\int\Sub{\MMod_\diamond}\MMod\prn{X}$ of a discrete fibration over
$\Tot{\MMod}\prn{\diamond}=\MMod_\diamond$. Here we are using the equivalence between $\DFib[\CC] \simeq \Psh{\CC}$. The preservation of
representable maps is then rendered here as the requirement that for
representable $\Mor[u]{X}{Y}$, each functor
$\Mor[\Tot{\MMod}\prn{u}]{\Tot{\MMod}\prn{X}}{\Tot{\MMod}\prn{Y}}$
shall have a right adjoint $\Tot{\MMod}\prn{u}\dashv
q\Sub{\Tot{\MMod}\prn{u}}$ taking an element of
$\Tot{\MMod}\prn{Y}$ to the \emph{generic} element of $\Tot{\MMod}\prn{X}$ in
the extended context.

\begin{example}
  For the representable map $\Mor[\pi]{\Tm}{\Ty}$, the functorial action
  $\Mor[\Tot{\MMod}\prn{\pi}]{\Tot{\MMod}\prn{\Tm}}{\Tot{\MMod}\prn{\Ty}}$ takes a term
  $\Gamma\vdash a:A$ to the type $\Gamma\vdash A$; the right adjoint
  $\Mor[q\Sub{\Tot{\MMod}\prn{\pi}}]{\Tot{\MMod}\prn{\Ty}}{\Tot{\MMod}\prn{\Tm}}$ sends a
  type $\Gamma\vdash A$ to the variable $\Gamma,a:A\vdash a:A$.
\end{example}

\begin{definition}\label{def:morphism-of-models}
  Given two models $\MMod,\NMod$ of $\TT$, a \DefEmph{morphism of
  models} from $\MMod$ to $\NMod$ is given by a natural
  transformation $\Mor[F]{\Tot{\MMod}}{\Tot{\NMod}} \in
  \brk{\TT^\rhd,\CAT}$ such that for each representable map $\Mor[u]{X}{Y}$ in $\TT$ the corresponding naturality datum $F_u\colon \Tot{\NMod}\prn{u}\circ F_X = F_Y\circ\Tot{\MMod}\prn{u}$ depicted below
  \[ 
    \begin{tikzpicture}[scale=0.5,baseline=($(nw)!0.5!(se)$)]
      \CreateRect{6}{4}
        \path
          coordinate[label=above:$\strut F_X$] (f0) at (spath cs:north 0.5)
          coordinate[label=above:$\strut \SmashTot{\NMod}\prn{u}$] (lB) at (spath cs:north 0.75)
          coordinate[label=below:$\strut \SmashTot{\MMod}\prn{u}$] (lA) at (spath cs:south 0.25)
          coordinate[label=below:$\strut F_Y$] (f1) at (spath cs:south 0.5)
        ;
        \draw[spath/save = f01] (f0) to (f1);
        \draw[spath/save = lBA] (lB) to[out=-90,in=90] (lA);
        \path[name intersections={of=f01 and lBA}]
          coordinate[dot,label=-10:$F_u$] (lambda) at (intersection-1)
        ;
        \begin{scope}[on background layer]
          \fill[catc] (nw) rectangle (se);
          \tikzset{
            spath/split at intersections={f01}{lBA},
            spath/get components of={lBA}\lcpts,
          }
          \fill[cate] (lB) to[spath/use={\getComponentOf\lcpts1,weld}] (lambda.center) to (f0) to cycle;
          \fill[catd] (lambda.center) to[spath/use={\getComponentOf\lcpts2,weld}] (lA) to (f1) to cycle;
          \fill[catf] (lB) to[spath/use={\getComponentOf\lcpts1,weld}] (lambda.center) to (f1) to (se) to (ne) to cycle;
        \end{scope}
    \end{tikzpicture}
  \] 
  satisfies the Beck--Chevalley condition in the sense the following 2-cell, obtained by conjugating with units and counits, denotes an invertible natural transformation $\Mor{F_X\circ
  q\Sub{\Tot{\MMod}\prn{u}}}{q\Sub{\Tot{\NMod}\prn{u}}\circ F_Y}$:
  \[
    \begin{tikzpicture}[scale=0.5,baseline=($(nw)!0.5!(se)$)]
      \CreateRect{6}{4}
      \path
        coordinate[label=above:$\strut F_X$] (f0) at (spath cs:north 0.5)
        coordinate[label=below:$\strut q\Sub{\Tot{\NMod}\prn{u}}$] (rB) at (spath cs:south 0.90)
        coordinate[label=above:$\strut q\Sub{\Tot{\MMod}\prn{u}}$] (rA) at (spath cs:north 0.1)
        coordinate[label=below:$\strut F_Y$] (f1) at (spath cs:south 0.5)
      ;
      \draw[spath/save = f01] (f0) to (f1);
      \path[spath/save = diagonal] (sw) to (ne);
      \path
        coordinate[dot,label=above:$\epsilon$] (counit) at (spath cs:diagonal 0.25)
        coordinate[dot,label=below:$\eta$] (unit) at (spath cs:diagonal 0.75)
      ;
      \draw[spath/save=swoosh] (rA) to[out=-90,in=180] (counit.center) to[out=0,in=180] (unit.center) to[out=0,in=90] (rB);
      \path[name intersections={of=swoosh and f01}]
        coordinate[dot,label=-10:$F_u$] (lambda) at (intersection-1)
      ;
      \begin{scope}[on background layer]
        \tikzset{
          spath/split at intersections={f01}{swoosh},
          spath/get components of={swoosh}\cpts,
        }
        \fill[catd] (nw) to (rA) to[spath/use={\getComponentOf\cpts1,weld}] (lambda.center) to (f1) to (sw) to (nw);
        \fill[catc] (rA) to[spath/use={\getComponentOf\cpts1,weld}] (lambda.center) to (f0) to cycle;
        \fill[cate] (f0) to (lambda.center) to[spath/use={\getComponentOf\cpts2,weld}] (rB) to (se) to (ne) to (f0);
        \fill[catf] (f1) to (lambda.center) to[spath/use={\getComponentOf\cpts2,weld}] (rB) to (f1);
      \end{scope}
    \end{tikzpicture}
  \]
\end{definition}

\begin{definition}
  Let $\MMod,\NMod$ be two models of $\TT$,
  and let $\Mor[F,G]{\MMod}{\NMod}$ be two morphisms of models.  An
  \DefEmph{isomorphism} $h$ from $F$ to $G$ is defined to be an \emph{invertible
  modification} between the underlying natural transformations $F,G$. This
  amounts to choosing for each $X\in\TT^\rhd$ a natural isomorphism
  $\Mor[h_X]{F_X}{G_X}$ in $\brk{\Tot{\MMod}\prn{X},\Tot{\NMod}\prn{X}}$, subject to the coherence condition that for each
  $\Mor[u]{X}{Y}$ in $\TT^\rhd$ the following two wiring diagrams are equal:
  \[
    \begin{tikzpicture}[scale=0.5,baseline=($(nw)!0.5!(se)$)]
      \CreateRect{6}{4}
        \path
          coordinate[label=above:$\strut F_X$] (f0) at (spath cs:north 0.5)
          coordinate[label=above:$\strut \SmashTot{\NMod}\prn{u}$] (lB) at (spath cs:north 0.75)
          coordinate[label=below:$\strut \SmashTot{\MMod}\prn{u}$] (lA) at (spath cs:south 0.25)
          coordinate[label=below:$\strut G_Y$] (f1) at (spath cs:south 0.5)
        ;
        \draw[spath/save = f01] (f0) to (f1);
        \draw[spath/save = lBA] (lB) to[out=-90,in=90] (lA);
        \path[name intersections={of=f01 and lBA}]
          coordinate[dot,label=-10:$G_u$] (lambda) at (intersection-1)
        ;
        \path (lambda.center) to coordinate[dot,label=left:$h_X$] (h/X) (f0);
        \begin{scope}[on background layer]
          \fill[catc] (nw) rectangle (se);
          \tikzset{
            spath/split at intersections={f01}{lBA},
            spath/get components of={lBA}\lcpts,
          }
          \fill[cate] (lB) to[spath/use={\getComponentOf\lcpts1,weld}] (lambda.center) to (f0) to cycle;
          \fill[catd] (lambda.center) to[spath/use={\getComponentOf\lcpts2,weld}] (lA) to (f1) to cycle;
          \fill[catf] (lB) to[spath/use={\getComponentOf\lcpts1,weld}] (lambda.center) to (f1) to (se) to (ne) to cycle;
        \end{scope}
    \end{tikzpicture}
    \quad
    \begin{tikzpicture}[scale=0.5,baseline=($(nw)!0.5!(se)$)]
      \CreateRect{6}{4}
      \path
        coordinate[label=above:$\strut F_X$] (f0) at (spath cs:north 0.5)
        coordinate[label=above:$\strut \SmashTot{\NMod}\prn{u}$] (lB) at (spath cs:north 0.75)
        coordinate[label=below:$\strut \SmashTot{\MMod}\prn{u}$] (lA) at (spath cs:south 0.25)
        coordinate[label=below:$\strut G_Y$] (f1) at (spath cs:south 0.5)
      ;
      \draw[spath/save = f01] (f0) to (f1);
      \draw[spath/save = lBA] (lB) to[out=-90,in=90] (lA);
      \path[name intersections={of=f01 and lBA}]
        coordinate[dot,label=100:$F_u$] (lambda) at (intersection-1)
      ;
      \path (lambda.center) to coordinate[dot,label=right:$h_Y$] (h/Y) (f1);
      \begin{scope}[on background layer]
        \fill[catc] (nw) rectangle (se);
        \tikzset{
          spath/split at intersections={f01}{lBA},
          spath/get components of={lBA}\lcpts,
        }
        \fill[cate] (lB) to[spath/use={\getComponentOf\lcpts1,weld}] (lambda.center) to (f0) to cycle;
        \fill[catd] (lambda.center) to[spath/use={\getComponentOf\lcpts2,weld}] (lA) to (f1) to cycle;
        \fill[catf] (lB) to[spath/use={\getComponentOf\lcpts1,weld}] (lambda.center) to (f1) to (se) to (ne) to cycle;
      \end{scope}
    \end{tikzpicture}
  \]
\end{definition}

\begin{remark}
  Because each of the induced maps
  $\Mor[\pi\Sub{\MMod\prn{X}}]{\Tot{\MMod}\prn{X}}{\MMod_\diamond}$ and
  $\Mor[\pi\Sub{\NMod\prn{X}}]{\Tot{\NMod}\prn{X}}{\NMod_\diamond}$ into the
  cone point are discrete fibrations, it suffices to check the modification
  condition of $h$ on only the cone maps $\Mor{X}{\diamond}$:
  as any discrete fibration is a faithful functor, it moreover follows that
  $\Mor[h_\diamond]{F_\diamond}{G_\diamond}$ uniquely determines all the
  other $h_X$ if they exist.
  Unfolding further, given $x\in \Tot{\MMod}\prn{X}$ we are only requiring that
  $\prn{h_\diamond}\Sub{\pi\Sub{\MMod\prn{X}}}^*\prn{G_X x} = F_X x$ in
  the sense depicted below in the discrete fibration $\NMod\prn{X}$ over $\NMod_\diamond$:
  \[
    \DiagramSquare{
      width = 3.5cm,
      nw/style = pullback,
      north/style = {exists,->},
      west/style = lies over,
      east/style = lies over,
      north = \exists! h_X x,
      nw = F_X x,
      ne = G_X x,
      sw = F_\diamond\prn{\pi\Sub{\MMod\prn{X}} x},
      se = G_\diamond\prn{\pi\Sub{\MMod\prn{X}} x},
      south = \prn{h_\diamond}\Sub{\pi\Sub{\MMod\prn{X}}}
    }
  \]
\end{remark}

Thus we have a (2,1)-category of models $\MOD{\TT}$ for any category with
representable maps $\TT$.

\section{The (2,1)-category of atomic substitution models}

\begin{definition}
  Given atomic substitution models $\Mor[\alpha]{\AtmMod}{\IMod}$ and
  $\Mor[\alpha']{\AtmMod'}{\IMod}$, a morphism from $\prn{\AtmMod,\alpha}$ to
  $\prn{\AtmMod',\alpha'}$ is given by a morphism
  $\Mor[F]{\AtmMod}{\AtmMod'}\in \MOD{\TT_0}$ together with an isomorphism
  $\Mor[\phi_F]{\alpha}{\alpha'\circ F}$ in $\brk{\AtmMod,\IMod}$ as depicted below:
  \[
    \begin{tikzpicture}[scale=0.5,baseline=($(nw)!0.5!(se)$)]
      \CreateRect{4}{3}
      \path
        coordinate[label=above:$\strut \alpha$] (alpha) at (spath cs:north 0.66)
        coordinate[label=below:$\strut \alpha'$] (alpha') at (sw -| alpha)
        coordinate[label=below:$\strut F$] (F) at (spath cs:south 0.33);
      \draw (alpha) to coordinate[dot,label=right:$\phi_F$] (phi/F) (alpha');
      \draw[spath/save=arc] (phi/F.center) to[out=190,in=90] (F);
      \begin{scope}[on background layer]
        \fill[catc] (nw) rectangle (se);
        \fill[catd] (F) to[spath/use={arc,reverse}] (alpha'.center) to cycle;
        \fill[catf] (alpha) rectangle (se);
      \end{scope}
    \end{tikzpicture}
  \]
\end{definition}

\begin{definition}
  Given two morphisms
  $\Mor[F,G]{\prn{\AtmMod,\alpha}}{\prn{\AtmMod',\alpha'}}$, an isomorphism
  from $F$ to $G$ is given by an isomorphism $\Mor[h]{F}{G}\in
  \brk{\AtmMod,\AtmMod'}$ such that the following wiring diagrams denote equal isomorphisms $\Mor{\alpha}{\alpha'\circ G}$:
  \[
    \begin{tikzpicture}[scale=0.5,baseline=($(nw)!0.5!(se)$)]
      \CreateRect{4}{4}
      \path
        coordinate[label=above:$\strut \alpha$] (alpha) at (spath cs:north 0.66)
        coordinate[label=below:$\strut \alpha'$] (alpha') at (sw -| alpha)
        coordinate[label=below:$\strut G$] (G) at (spath cs:south 0.33);
      \draw (alpha) to coordinate[dot,label=right:$\phi_F$] (phi/F) (alpha');
      \draw[spath/save=arc] (phi/F.center) to[out=190,in=90] coordinate[dot,label=left:$h$] (h) (G);
      \begin{scope}[on background layer]
        \fill[catc] (nw) rectangle (se);
        \fill[catd] (G) to[spath/use={arc,reverse}] (alpha'.center) to cycle;
        \fill[catf] (alpha) rectangle (se);
      \end{scope}
    \end{tikzpicture}
    \qquad
    \begin{tikzpicture}[scale=0.5,baseline=($(nw)!0.5!(se)$)]
      \CreateRect{4}{4}
      \path
        coordinate[label=above:$\strut \alpha$] (alpha) at (spath cs:north 0.66)
        coordinate[label=below:$\strut \alpha'$] (alpha') at (sw -| alpha)
        coordinate[label=below:$\strut G$] (G) at (spath cs:south 0.33);
      \draw (alpha) to coordinate[dot,label=right:$\phi_G$] (phi/G) (alpha');
      \draw[spath/save=arc] (phi/G.center) to[out=190,in=90] (G);
      \begin{scope}[on background layer]
        \fill[catc] (nw) rectangle (se);
        \fill[catd] (G) to[spath/use={arc,reverse}] (alpha'.center) to cycle;
        \fill[catf] (alpha) rectangle (se);
      \end{scope}
    \end{tikzpicture}
  \]
\end{definition}

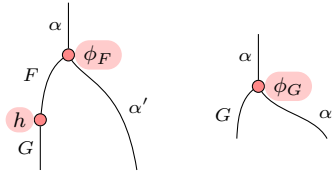
\end{document}